\documentclass[titlepage,12pt]{article}
\pdfoutput=1
\usepackage{fancyhdr}
\usepackage{lastpage}
\usepackage[utf8]{inputenc} 
\usepackage{amsmath,amssymb,amsfonts,mathrsfs,latexsym}
\usepackage{ulem}
\usepackage{tabularx}
\usepackage{threeparttable}
\usepackage{threeparttablex}
\usepackage{booktabs}
\usepackage{setspace} 
\usepackage[tablename=TABLE,labelsep=newline,aboveskip=0pt,bf,justification = centering]{caption}
\usepackage[dvipsnames]{xcolor}
\usepackage[capposition=top]{floatrow}
\usepackage{graphicx}
\usepackage{titlepic}
\usepackage{enumerate}
\usepackage{enumitem}
\usepackage{subcaption}
\usepackage{floatrow}
\usepackage{caption}
\usepackage[colorlinks = true,
            linkcolor = blue,
            urlcolor  = blue,
            citecolor = blue,
            anchorcolor = blue]{hyperref}

\usepackage[margin=2.5 cm,includehead,includefoot]{geometry} 
\usepackage{array}
\newcolumntype{P}[1]{>{\centering\arraybackslash}p{#1}}
\usepackage{subcaption}

\usepackage{changepage}

\usepackage{titlesec}
\titleformat*{\section}{\large\bfseries}
\titleformat*{\subsection}{\normalsize\bfseries}
\titleformat*{\subsubsection}{\small\bfseries}

\usepackage[backend=biber, citestyle=authoryear-comp,natbib]{biblatex} 
\usepackage[nottoc,numbib,numindex,section]{tocbibind}
\usepackage{hyperref}
\hypersetup{
    colorlinks=false
}

\numberwithin{equation}{section}

\newgeometry{left=2.5cm,right=2.5cm,top=2.5cm,bottom=2cm}
\begin{document}
\thispagestyle{empty}
\singlespacing
\title{Understanding the Distributional Aspects of Microcredit Expansions}
\author{Tobias Gabel Christiansen \\
Faculty of Economics \smallskip \\ University of Cambridge \smallskip \\
\and Melvyn Weeks\thanks{\small{Contact Author: Dr. M. Weeks, Faculty of Economics,
University of Cambridge, Cambridge
CB3 9DD, UK. Email: mw217@econ.cam.ac.uk} } \\
Faculty of Economics and Clare College \smallskip \\ University of Cambridge}
{\let\newpage\relax\maketitle}
\maketitle
\begin{minipage}{0.9\linewidth}
\hspace{0.75cm}
\hangindent = 1.5cm 
\small{\hspace{0.6cm}\textbf{Abstract.} Various poverty reduction strategies are being implemented in the pursuit of eliminating extreme poverty. One such strategy is increased access to microcredit in poor areas around the world. Microcredit, typically defined as the supply of small loans to underserved entrepreneurs that originally aimed at displacing expensive local money-lenders, has been both praised and criticized as a development tool (\cite{RCTeval}). 
This paper presents an analysis of heterogeneous impacts from increased access to microcredit using data from three randomised trials. In the spirit of recognising that in general the impact of a policy intervention  varies conditional on an unknown set of factors, particular, we investigate whether heterogeneity presents itself as groups of winners and losers, and whether such subgroups share characteristics across \textsc{rct}s. We find no evidence of impacts, neither average nor distributional, from increased access to microcredit on consumption levels. In contrast, the lack of average effects on profits seems to mask heterogeneous impacts. The findings are, however, not robust to the specific machine learning algorithm applied. Switching from the better performing Elastic Net to the worse performing Random Forest leads to a sharp increase in the variance of the estimates.  In this context, methods to evaluate the relative performing machine learning algorithm developed by \citet{CDDF} provide a disciplined way for the analyst to counter the uncertainty as to which algorithm to deploy. \\


 \hspace{1.5 cm}JEL Classification Codes: D14, G21, I38, O12, O16, P36 \\

\hangindent = 1.5cm 
\hspace{1.5cm}Keywords: Machine learning methods, microcredit, development policy, treatment effects, random forest, elastic net\\

\vspace{1cm}

}

\end{minipage}

\section{Introduction}
According to the World Bank there has been marked progress in reducing poverty over the past decades, and the number of people living at or below the poverty line in 2015 was close to one third of the number in 1990. Yet, the number of people currently living in extreme poverty still amounts to 10 percent of the world’s population  (\cite{WBintro}). Various poverty reduction strategies have, and are, being implemented in the pursuit of eliminating extreme poverty. One such strategy is increased access to microcredit in poor areas around the world. Microcredit, typically defined as the supply of small loans to underserved entrepreneurs that originally aimed at displacing expensive local money-lenders, has been both praised and criticized as a development tool  (\cite{RCTeval}). In 2006, M. Yunus received the Nobel Prize for pioneering the concepts of microcredit as a measure to combat poverty by allowing poor entrepreneurs to expand their businesses and thereby generate economic prosperity  (\cite{Yunus}). More recently, A. Banerjee, E. Duflo, and M. Kremer jointly received the Nobel Prize in 2019 for their experimental approach to alleviating global poverty, which includes experiments concerning microcredit  (\cite{DufloIntro}). Others blame microcredit for harming the poor by creating over-indebtedness, and displacing existing business owners by increasing competition and eating local demand (\cite{MicroCritique}). \\

Both the positive and negative claims may carry some truth given that initiatives, such as increased access to microcredit, rarely have homogenous effects leaving everyone better or worse off.
Distributional treatment effects are therefore important
as the impact of microcredit is likely to be heterogenous across individuals (see, for example, \citet{MeagerAGG}). Understanding the heterogeneity is of great interest to policy makers as it illuminates how treatment affects sub-groups of the population differently. Solely focusing on average effects may be of little use if some groups benefit,
while other lose, thereby creating a zero net effect on average when considering the entire population. 
A large number of randomised evaluations of microfinance institutions (\textsc{mfi}s) have been deployed around the world and yet the consensus about the overall effect of microcredit has been questioned due to concerns about the external validity of the findings across \textsc{rct}s  (\cite{RCTeval,contextmatters}). Learning the characteristics of potential winners and losers across \textsc{rct}s may be of use in predicting the effect of microcredit expansions in future settings that consider different populations with different characteristics (\cite{RmeetsM}). \\ 

This paper presents an analysis  of heterogenous impacts of increased access to microcredit, using data from three \textsc{rct}s.
In particular, we study heterogeneous impacts on monthly business profits and monthly consumption using data from three studies that all considered increased access to microcredit as the main intervention, employed random assignment of the access to microcredit,
and  collected data on large rich number of baseline covariates, thereby enabling the exploration of heterogeneity in the treatment effects. The studies we consider are \citet{Morocco}, \citet{Mongolia}, and \citet{Bosnia} which were originally deployed in Morocco, Mongolia and Bosnia \& Herzegovina, respectively. We focus on monthly business profits to evaluate the claim that microcredit allow poor entrepreneurs to expand their businesses and increase profits.\footnote{Another claim is that microcredit allow the poor to open businesses  \citet{MeagerAGG}. The effect of entry is implicitly captured by profits as new entrants will affect the estimates.} Moreover, as households may benefit from microcredit in other ways, we also study heterogeneous effects on total monthly consumption which can be viewed as a crude proxy for welfare benefits.\\

A key contribution of this paper to the microcredit literature  is the use of machine learning as a tool to uncover differences in outcomes. Knowledge of how microcredit affects profits and consumption differently across households is valuable in deciding whether microcredit is worth allocating resources towards, and in particular, inform policy makers about the presence of potential winners and losers. Specifically, the use of machine learning helps in overcoming traditional issues related to multiple hypothesis testing, by providing an agnostic approach to defining subgroups with different treatment effects, in conjunction with locating the characteristics of these subgroups that are the most important in predicting the heterogeneity. The method improves on alternative ways of dealing with concerns of multiple hypothesis, such as pre-analysis plans, by allowing researchers to take full advantage of the rich baseline datasets, that are often collected in microcredit \textsc{rct}s.\\

In addition, the method is generic with respect to the machine learning algorithms used to estimate the heterogeneity in treatment effects, as it works under biased and even inconsistent machine learning estimators. To take advantage of this,
two machine learning algorithms with different strengths and weaknesses are applied. Using an Elastic Net regression and a Random Forest we analyse
whether heterogeneity in treatment effects exist across studies and whether it presents itself as subgroups of winners and losers.
Across two of the three \textsc{rct}s we find evidence of heterogeneous effects on monthly profits, which presents itself as groups of winners and losers, thereby supporting both the negative and positive claims of microcredit. Moreover, the heterogeneity in treatment effects can be predicted from household level covariates, and the most affected households are less indebted, engage in self-employment activities, and consume less at baseline. The findings are, however, not robust to a change in the machine learning algorithm used in conjunction with the method developed by \citet{CDDF}. Turning to monthly consumption, there is no evidence of average nor distributional effects from increased access to microcredit. \\

The paper proceeds as follows. Section 2 gives a brief review of the microcredit literature and the studies analysed in this paper. Section 3 provides a thorough outline of the method applied. Section 4 presents the results and section 5 concludes.

\section{Microcredit Expansions: Experiments and Data}
In the past decade numerous \textsc{rct}s have been implemented in a variety of settings in the pursuit of understanding the effects of microcredit (see \citet{Morocco,Mongolia,Bosnia,india,Philli,Ethiopia, Mexico}).  The main focus has been on average effects on a myriad of different outcome variables, spanning from profits and consumption to educational attainment and women's empowerment. Even though most studies find some evidence towards increased take-up rates of microloans from treatment, there is a lack of evidence of significant effects on the average household. Most studies therefore find little or no support for neither the positive nor the negative claims of the effects of microcredit. Some of the studies also present a limited analysis of heterogeneous effects, primarily by estimating quantile treatment effects.  Other methods such as sub-sample analysis or multiplicative models based on interactions between covariates and the treatment dummy are also provided. In addition, \citet{contextmatters} conduct an meta-analysis of six experiments and concludes that attention should be pointed towards heterogeneous treatment effects as little can be learned from analyses that focus exclusively on, for the most part insignificant, estimates of average effects.\\


\citet{meagerUNDER,MeagerAGG} also conduct a meta-analysis in which she aggregates the data from the seven \textsc{rct}s above  using a Bayesian Hierarchical model.  
In particular, \citet{MeagerAGG} analyses heterogeneous impacts of microcredit using quantile treatment effects and finds that microcredit has negligible impacts on profits, revenues, expenses and consumption for households below the 75th percentile across settings. \citet{meagerUNDER} focuses upon average treatment effects and finds that average effects on profits, revenues, expenses and different measures of consumption from increased access to microcredit may be limited. Turning to heterogeneity, she finds that approximately 60 percent of the observed variation in treatment effects across studies is due to sampling variation as opposed to genuine heterogeneity in effects. In an attempt to explain  the variation in treatment effects across studies she interacts the treatment indicator with previous business experience and finds that existing business owners benefit the most from increased access to microcredit.
That previous business experience is important in predicting heterogeneity in treatment effects is also found by \citet{hettreat}.



\subsection{The Randomised Experiments}
Table \ref{t1} summarizes the three studies analysed in this paper. The studies all meet the following inclusion criteria: the main intervention studied is an expansion of microcredit access, the assignment of access is randomised, and data includes a large  number of baseline covariates which can be used to uncover heterogeneity in treatment effects.\footnote{The studies were published in a special edition of \textsc{aej}: Applied Economics. The original articles and datasets are available at \url{https://www.aeaweb.org/issues/360}.}  Apart from these requirements the studies differ across a range of factors. In particular, the studies were deployed across three different countries (Morocco, Mongolia and Bosnia \& Herzegovina). In addition, and as detailed in Table \ref{t1}, they differ in their treatment methods, sampling frames, eligibility criteria and microloan liabilities and therefore target different populations. Comparing treatment effects from studies that differ to such an extent, while still maintaining the same overall goal, has the potential to reveal commonalities in the factors driving the effects from increased access to microcredit. 
To the extent that heterogeneous treatment effects are manifest as subgroups of winners and losers that share similar characteristics, is useful in identifying mechanisms that underpin these differences.\\


We study the effect of randomly assigned increase in \textit{access} to credit from \textsc{mfi}s on monthly business profits and total monthly consumption, both measured in \textsc{usd (\textsc{ppp})} indexed to 2009.\footnote{Data on the \textsc{ppp} conversion factors can be found at the World Bank Open Data (ID: PA.NUS.PRVT.PP).} The increase in access to microfinance occurred through \textsc{mfi}s randomly opening branches at the community level in the \textsc{rct}s in Morocco and Mongolia, and from random offers made at the individual level in Bosnia \& Herzegovina. Importantly, the main parameter of interest is not actual loan take up but increased access to microcredit which may be interpreted as the intent-to-treat effect. The reason for this is that the stable unit treatment value assumptions (\textsc{sutva}) are likely to fail due to informal financial links between households and general equilibrium effects in the \textsc{rct}s where randomisation were done a the community level. Indeed, \citet{india} argue that the possibility of general equilibrium effects or spillovers, such as effects on prices and wages,
are likely. Moreover, as pointed out by \citet{Mongolia}, focusing on the effects from access to credit  allows a policy maker to learn about the effects on the population initially targeted and not just those who take up loans.\\   



An important feature of the three studies considered in this paper is the collection of pre-treatment characteristics which can be used to uncover heterogeneity in treatment effects, by defining subgroups of the population under consideration. Importantly, characteristics used to create subgroups must be immutable, such as age or gender, or recorded at baseline (before randomisation) to rule out the possibility that they are affected by treatment, and thereby potential mediators distorting treatment effect estimates (see  \citet{controls}). 
Tables \ref{Morocco}-\ref{Bosnia} show the covariates used in our analysis along with comparisons of the mean values for treated and untreated. Notably, for the study in Morocco an additional 1433 observations were randomly sampled at the endline, and no baseline covariates are available for these individuals. As these observations are missing completely at random we include them by setting the missing values equal to a constant (zero) and adding a dummy variable 
indicating whether the observation is missing or not.\footnote{In addition to the random missingness, there is a small number of likely non-random missing observations across a large number of covariates for the study in Morocco. We drop these observations, which amounts to approximately 3 percent of the sample, reducing the overall sample size to 5329 observations. Table \ref{Morocco} suggests that the covariates remain balanced across treated and untreated after doing so.}\\

In contrast to \citet{Mongolia} who solely focus on people in group treatment (i.e. joint liability lending)  we  include people from both group and individual treatment schemes to enrich the sample. We then add a dummy indicating whether an individual was assigned to either type of treatment. In addition, and in contrast to \citet{Morocco}, we abstain from trimming the sample to only include people with a high probability of borrowing as our focus is on \textit{access} to microcredit rather than actual loan take up. Apart from these changes  we conform to the decisions made by the original authors regarding the analysis and definitions of the outcome variables throughout the paper.

\newpage
\newgeometry{left=2.3cm, right=2cm}
\begin{center}
\captionof{table}{Summary of the Randomised Control Trials}\label{t1}
\scalebox{0.75}{
\begin{tabular}{lP{5cm}P{5cm}P{5cm}}
\hline\hline
Country & Morocco & Mongolia & Bosnia \& Herzegovina \\ \hline
Study citation & Crépon et al. (2015) & Attanasio et al. (2015) & Augsburg et al. (2015) \\
 Treatment & Open branches  & Open branches target likely borrowers & Lend to marginally rejected borrowers \\
 Randomisation level & Community (162 villages)  & Community (40 villages) & Individual\\
 Urban or Rural & Rural  & Rural & Both\\
 Target women? & No & Yes & No\\
 Target microentrepreneurs? & Yes & Yes & Yes\\
 MFI already operating& No & No & Yes\\
 Microloan liability type & Group (3-4 people) & Both group and individual (7-15 people in groups) & Individual \\
 Collateralized & No & Yes & Yes\\
 Loan take up in treatment & 17 pct. & 54 pct.  & 100 pct.\\
 Average loan size per borrower & \$1374  & \$588  & \$1848 \\
 Interest rate$^{a)}$& 14.5 percent APR & 26.8 percent APR  & 22 percent APR \\
 Sampling frame & Random sample plus likely  borrowers & Women who registered interest in loans and met eligibility criteria. & Marginal applicants considered too risky to be offered credit as regular borrowers.\\
 Loan eligibility & Men and women ages 18-70 who holds a national ID card, a residency certificate and have had an economic activity other than non-livestock agriculture for at least 12 months. & Women who own less that MNT 1 million (\$1948) in assets and earn less than MNT 0.2 million (\$389) in monthly profits from a business. & Men and women with sufficient collateral, repayment capacity, credit worthiness. business capacity etc.\\
 Response rate at endline  & 92 pct. & 84 pct. & 83 pct. \\
 Sample size$^{b)}$  & 5524 & 960 & 995 \\
 Baseline covariates exist & Yes$^{c)}$ &  Yes  & Yes\\
 Study starting date & April 2006 & March 2008 & January 2009 \\
 Study duration & 24 months & 19 months  & 14 months\\ \hline \hline
\end{tabular}}
\begin{tablenotes}[para,flushleft]
\scriptsize	
 Notes$:$ The table presents characteristics of the original studies. All dollar (\$) values are \textsc{usd (ppp)} indexed to 2009.\newline
 $^{a)}$Annual Percentage Rate (\textsc{apr}) given by the upper bound of the interest rate ranges reported for each study. \newline
 $^{b)}$Sample sizes are defined as the endline sample sizes i.e. after attrition has been accounted for.\newline
 $^{c)}$For the \textsc{rct} in Morocco an additional 1433 individuals were randomly sampled at the endline to increase the sample size. For these additional individuals no baseline covariates exist.\newline
\end{tablenotes}
\onehalfspacing
\end{center}
\restoregeometry
\newpage



\section{Estimating Heterogenous Treatment Effects}
The strategy employed in this paper is a method recently developed by \citet{CDDF} for estimating heterogenous treatment effects in RCTs using machine learning. The method builds upon the potential outcomes framework introduced by \citet{Neyman}, and extended by \citet{Rubin}, and defines the main causal functions as the Baseline Conditional Average (BCA),
\begin{align}
    b_0(Z) \equiv \mathbb{E}[Y(0)|Z] \label{(2.1)}
\end{align}
and the Conditional Average Treatment Effect (CATE),
\begin{align}
    s_0(Z) \equiv \mathbb{E}[Y(1)|Z]-\mathbb{E}[Y(0)|Z] \label{(2.2)}
\end{align}
where $Y(1)$ and $Y(0)$ are the potential outcomes in the treatment state ($D=1$) and non-treatment state ($D=0$), defined by the binary treatment indicator $D\in\{1,0\}$, $Z$ is a vector of covariates that characterise the observational unit, and $\mathbb{E}[\cdot]$ is an expectation operator.\footnote{$D$ indicates whether or not an individual experiences increased \textit{access} to microcredit.} The data observed is $(Y_i, Z_i, D_i)^N_{i=1}$, consisting of $N$ i.i.d. draws of the random vector $(Y,Z,D)$. In this framework $s_0(Z)$ can be used to discover heterogeneity in treatment effects at the individual level or for groups by comparing outcomes between subgroups defined by $Z$. However, as is well known in the causal inference literature, and was labeled the fundamental problem of causal inference by \citet{Holland}, the key issue to estimating $s_0(Z)$ is the fact that either $Y(1)$ or $Y(0)$ is observed for each observational unit, but never both. In order to identify $s_0(Z)$
the following assumptions of unconfoundedness and overlap are made. \\

\textit{Unconfoundedness}\\
Unconfoundedness requires that all confounding information for the relation between the treatment and the potential outcomes is captured by the observed covariates $Z$. In other words, $D$ has to be as good as random conditional on $Z$. Mathematically, it can be expressed as,
\begin{align}
    Y(1),Y(0) \perp D | Z \label{2.3}
\end{align}
where $\perp$ denotes independence. \\

\textit{Overlap}\\ 
This requires any observation to have a positive probability of being assigned to either the treatment or control group.  Alternatively put, there has to be individuals who are treated and untreated for all possible values of $Z$,
\begin{align}
    0<\mathbb{P}(D=1|Z)=\mathbb{E}(D=1|Z)=p(Z)<1 \quad \forall z \in Z \label{overlap}
\end{align}
where $p(Z)$ denotes the propensity score which measures the probability that an individual receives treatment conditional on covariates.
These assumptions together are referred to as the assumption of strong ignorability in the paper by \citet{RR}.
Both assumptions are naturally satisfied in a randomised experiment given proper randomisation as this ensures that treated and untreated individuals are comparable (in expectation) across both observed and unobserved covariates, with the only difference being the treatment. This makes it possible to interpret differences in outcomes $Y$ across treated and untreated, conditional on $Z$, as the causal effect of treatment.\footnote{Given proper randomisation it is not necessary to condition on $Z$ for (\ref{2.3}) to hold. However, in practice, and as discussed by \citet{Deaton1} and \citet{Deaton2}, some variables may be imbalanced in \textit{finite} RCTs. Controlling for $Z$ can therefore help with avoiding bias, as well as improve on the precision of the estimated treatment effect.} Under the assumption of strong ignorability, the causal functions in (\ref{(2.1)}) and (\ref{(2.2)}) are identified.

\subsection{Linear Framework}
 An estimate of the \textsc{cate} can be obtained
by including interactions of the treatment indicator and the covariates of interest in a linear interaction model. Letting $X_i$ be a single observed covariate then one model of the \textsc{cate} is given by,
\begin{align}
    Y_i = \rho_0 + \rho_1 D_i+\rho_2X_i+ \rho_3(D_i\times X_i)+\nu_i \label{linCATE}
\end{align}

In (\ref{linCATE}) $\rho_1$ is the ATE and $\rho_1+\rho_3\times X_i$ is a linear approximation of the \textsc{cate} which measures how the marginal effect of treatment depends on the value of $X_i$  \citet{Gell}.
The estimate of the \textsc{cate} obtained from (\ref{linCATE}) is a linear function of $X_i$ by construction. Moreover, the interaction involves only a single covariate. This is often the case in practice where $X_i$ equals things such as gender, age, or previous business experience, thereby limiting the analysis of heterogeneous effects to a single dimension of the covariate space at a time. Examples from the microcredit literature include \citet{Philli} who analyse heterogenous treatment effects of multiple outcomes, including business size, by interacting a loan assignment indicator and a gender dummy in an OLS setting, and \citet{Ethiopia} who analyse heterogenous impacts on labor supply by restricting the sample to different age groups before applying OLS.\footnote{Most studies in the microcredit literature also include an analysis of heterogenous impacts by estimating quantile treatment effects (\cite{india}, \cite{Morocco},\cite{Ethiopia} and others)}\\

An issue with this approach is that researchers often have little knowledge, a priori, about which dimensions of the covariate space $Z$ that are the most relevant in detecting heterogeneity in treatment effects. Moreover, the true functional form of $s_0(Z)$ is rarely known, and hence the standard linear model may not be the most suitable for accurately depicting the heterogeneity in treatment effects  (\cite{IW}). A concern of multiple hypothesis testing and invalid inference then arises if researchers search over the covariate space and alternative model specifications until they find heterogeneity of interest and reports these findings as confirmatory results, without giving an external party the ability to check the validity. The issue is that the size of the tests in such scenarios are invalid, and hence the probability of falsely rejecting a null hypothesis is larger than what it appears to be (\cite{wallach}).
Methods that adjust the sizes of the test, such as the Bonferroni and the Hochbergs corrections (\cite{Bonferroni,Hochberg}), only correct the sizes of the test statistics presented. If more tests were conducted but not presented, the adjustment fails and inference becomes invalid.\\

One way to overcome such concerns is to introduce pre-analysis plans
which consists of a report prior to the analysis stating the covariates to be used in the analysis of heterogeneity in treatment effects. This makes it clear what additional tests were run beyond those
originally planned, and thus it makes multiple testing adjustment more credible (\cite{MeagerBITSS}).
Some, including \citet{AtheyTalk}, \citet{CDDF} and \citet{Deaton2}, have however pointed out that pre-analysis plans amount to throwing away lots of valuable information by restricting heterogeneity analyses to pre-registered subgroups. This has given rise for other techniques to gain ground, some of which apply machine learning.

\subsection{The Machine Learning Approach}

Machine learning is a broad term that covers a variety of different statistical learning methods, including Random Forests, Elastic Net Regularization, Neural Networks and others. Machine learning is centered around prediction and distinguishes itself from the traditional frequentist approach by using sample splitting for model selection, estimation and testing the model of interest. This is inherently different from traditional econometrics which uses all the available data to estimate the parameters of a pre-specified model and aims at drawing causal inference (\cite{AtheyTalk}).
Machine learning has proven to be excellent at handling high-dimensional data as it is effective at exploring various functional forms and discovering the most relevant covariates in order to yield accurate predictions. \\

Redirecting the goal of machine learning from prediction to causal inference allows researchers to reap these benefits. In particular, it allows researchers to take an agnostic and hypothesis-free approach in discovering heterogeneity in treatment effects across subgroups of the population in high-dimensional settings, by letting the data drive the model and covariate selection (\cite{AtheyHandbook,CDDF}). In particular, it is beneficial in relation to the microcredit literature as it allows researchers to take full advantage of the rich baseline characteristics that are often collected during randomised experiments.\\

Ideally, machine learning 
would therefore be used to estimate general functional forms of the \textsc{bca}. Under the assumption of strong ignorability the general functional forms of the \textsc{bca} and \textsc{cate} can be identified as,
\begin{align}
    b_0(Z)=\mathbb{E}[Y|D=0,Z]=f_b(Y,Z,D) \label{b0}
\end{align}
and
\begin{align}
    s_0(Z)=\mathbb{E}[Y|D=1,Z]-\mathbb{E}[Y|D=0,Z]=f_s(Y,Z,D) \label{s0}
\end{align}
Multiple different machine learning algorithms could be used to provide estimators for $b_0(Z)$ and $s_0(Z)$. The estimators of $b_0(Z)$ and $s_0(Z)$ from naively applying machine learning are, however, likely to be both biased and inconsistent.

\subsubsection{Naive Machine Learning Estimator of the CATE}
A naive way to obtain estimators of $b_0(Z)$ and $s_0(Z)$ (i.e. $B(Z)$ and $S(Z)$) using machine learning, involves splitting the sample into two, with one part containing all the treated and the other all the untreated individuals. $B(Z)$ is then given by the estimator obtained from fitting a given machine learning algorithm (e.g. Elastic Net, Random Forest) to the subsample with the untreated individuals, and $S(Z)$ is given by the difference between the two estimators obtained from fitting the machine learning method to each subsample. The estimators obtained in this way are naive in the sense that they apply methods developed for prediction in an attempt to draw inferences about model parameters (treatment effects) without accounting for the model selection (regularization) that comprises the machine learning algorithms (\cite{Belloni}). In particular, most machine learning methods
fail to provide valid inference
as they
fail to yield consistent estimators of $s_0(Z)$ in high-dimensional settings without strict, and often not testable assumptions, such as assumptions about the sparsity of the relation between 
the outcome and the covariates (\cite{CDDF}). The issue with machine learning estimators of conditional treatment effects stems from biases that result from model selection (i.e. regularization).\footnote{An exception is the Causal Forest algorithm developed by \citet{cf2} which is an adaptation of the Random Forest that aims at providing valid pointwise inference for the \textsc{cate}. However, as \citet{CDDF} point out the consistency of the Causal Forest hinges on the dimension of $Z$ being less than $\log(n)$. Causal Forests are therefore only guaranteed to provide valid results in low-dimensional settings which contrasts the settings considered in this paper.}\\

\textit{Regularization Bias}\\
Most machine learning methods have 
prediction at their core and use the Mean Squared Error (\textsc{mse}) as the objective to minimize. This minimization often relies on regularization of the machine learning methods to avoid overfitting the data used to estimate the model and thereby keeping the variance of the resulting estimator from exploding. It does, however, also induce a bias in $S(Z)$. The issue with the naive machine learning estimators described above arises as the treatment indicator is implicitly forced to remain in the models (as it defines the subsamples) and hence $D$ becomes immune to regularization. Any variable that is highly correlated with the treatment variable therefore 
tends to be excluded or neglected from the model since the variable does not add much predictive power for the outcome given the treatment indicator is already controlled for. If such variables also correlate with the outcome then regularization is effectively excluding confounding variables from the model leading to a biased estimator of $s_0(Z)$  (\cite{Belloni}).
Counter to what is commonly presumed in \textsc{rct}s, the presence of confounding variables in \textsc{rct}s may indeed be an issue as the likelihood that some variables are imbalanced across treated and untreated, and thereby correlated with $D$, is high in \textit{finite} \textsc{rct}s (\cite{Deaton1,Deaton2}).\footnote{By inspection of Tables \ref{Morocco}-\ref{Bosnia} it is evident that a few variables in each \textsc{rct} are imbalanced and correlates with treatment.} The regularization bias from neglecting confounders will in general converge to zero at very slow rates, thereby invalidating inference that relies on assumptions of asymptotic normality (\cite{DDML}).

\subsection{Machine Learning Proxies}
The method developed by \citet{CDDF} allows for estimating heterogeneous treatment effects in randomised experiments using machine learning which is valid in high-dimensional settings. In particular, the method avoids imposing strong conditions on the naive machine learning estimators as it works under biased and even inconsistent learning. Instead, \citet{CDDF} simply treat the naive machine learning estimators described in the previous subsection as proxies that serve in place of $b_0(Z)$ and $s_0(Z)$. The machine learning proxy of $s_0(Z)$ is then used to create an orthogonalized variable which together with an orthogonalized treatment indicator identify features or objects of the \textsc{cate} function in separate regression models. The method is therefore a two-step procedure in which machine learning is used as a part of a less ambitious goal that aims at describing a few properties of the \textsc{cate} function (\ref{s0}) rather than the entire \textsc{cate} function. \\

\citet{CDDF} propose the following steps to obtain the machine learning proxies,
\begin{enumerate}
  \item Randomly split one of the RCT samples into a main sample $(Data_M)$ and a auxiliary sample $(Data_A)$ of equal size.
  \item Use $Data_A$ to estimate a machine learning algorithm (e.g. Elastic Net, Random Forest) to predict $Y$ using $Z$. In particular, $Data_A$ is split into treated and untreated individuals and the machine learning method is fit to each subsample.
  \item For each $i\in Data_M$ predict $Y_i$ for each observation in $Data_M$ using both prediction models from the previous step to obtain two predicted outcomes, $\hat{Y}_i^{D=1}$ and $\hat{Y}_i^{D=0}$.
  \item Calculate the proxy of $b_0(Z)$ using the model trained on the control group $B(Z)=\hat{Y}_i^{D=0}$ and the proxy of $s_0(Z)$ as the difference between the predictions $S(Z)=\hat{Y}_i^{D=1}-\hat{Y}_i^{D=0}$.
\end{enumerate}
The procedure is done twice with $Y$ equal to monthly profits and monthly consumption for each RCT, and two pair of the proxies $B(Z)$ and $S(Z)$ are obtained. In both cases, $Z$ is given by the covariates presented in Tables \ref{Morocco}-\ref{Bosnia}.\footnote{For the RCTs in Morocco and Mongolia, we follow the advice by \citet{CDDF} and use stratified sample splitting.} The proxies obtained from the above procedure are similar to the naive machine learning estimators described, and $S(Z)$ is likely to be a biased and inconsistent estimator of $s_0(Z)$ due to regularization. Nevertheless, \citet{CDDF} show that $S(Z)$ can still be used to extract important \textit{features} of $s_0(Z)$.

\subsubsection{Key Features of the CATE}\label{features}
\textit{Best Linear Predictor}\\
The first feature is the Best Linear Predictor (\textsc{blp}) of the \textsc{cate} function based on the proxy $S(Z)$,
\begin{align}
    \beta_1 + \beta_2 (S(Z)-\mathbb{E}[S(Z)]) \label{lin}
\end{align}
Here, $\beta_1$ is the average 
treatment effect (ATE),
\begin{align}
    \beta_1 = \mathbb{E}[s_0(Z)]
\end{align}
and $\beta_2$ captures any additional heterogeneity in treatment effects. The \textsc{blp} of the \textsc{cate} therefore provides individualized treatment effects. \citet{CDDF} show that, 
\begin{align}
    \beta_2 = \frac{\text{Cov}(s_0(Z),S(Z))}{\text{Var}(S(Z))} \label{b2}
\end{align}
From (\ref{b2}) it is evident that $\beta_2$ captures how well the proxy $S(Z)$ approximates $s_0(Z)$. If $S(Z)$ is a perfect proxy then $\beta_2=1$ and if $S(Z)$ is complete noise then $\beta_2=0$. In addition, if there is no heterogeneity in the treatment effects, such that $s_0(Z)=s_0 \in \mathbb{R}$, then $\beta_2=0$. Rejection of the null hypothesis of $\beta_2=0$ therefore implies that heterogeneous treatment effects exist and $S(Z)$ is a effective predictor of it.\\

Empirically, the coefficients of the \textsc{blp} can be estimated  from a weighted least squares (\textsc{wls}) regression using the main sample and the proxies $B(Z)$ and $S(Z)$,
\small
\begin{align}
    Y_{ij} = \alpha_{0}+ \omega_j + \alpha_{1} B(Z_{i})+\alpha_{2} S(Z_{i})+ \beta_{1}(D_{i}-p(Z_{i}))+\beta_{2}(D_{i}-p(Z_{i})(S(Z_{i})-\overline{S(Z)})+\epsilon_{{ij}} \label{BLP}
\end{align}
\normalsize
Here, the subscript $i$ refers to the individual and $j$ to the village/province. The weights used are the inverse of the variance of $D_i$ i.e. $w(Z_{i})=(p(Z_{i})(1-p(Z_{i}))^{-1}$ where $p(Z_{i})$ denotes the propensity score. The estimator of $p(Z_i)$ is set to the proportion of treated individuals in each \textsc{rct} because treatment was randomly assigned. $\overline{S(Z)}=|M|^{-1}\sum_{i\in M}S(Z_{i})$ is the empirical expectation of $S(Z_i)$ with respect to the main sample. Equation (\ref{BLP}) is estimated for both monthly profits and monthly consumption, with clustered standard errors at the village level for the \textsc{rct}s in Morocco and Mongolia to account for the possibility of correlated shocks within villages. For the \textsc{rct} in Bosnia \& Herzegovina standard errors robust to heteroskedasticty are used. $\omega_j$ are strata dummies at the village level for the \textsc{rct} in Morocco, and at the province level for the \textsc{rct} in Mongolia. No strata dummies are included for the RCT in Bosnia \& Herzegovina as randomisation was not stratified on regions.\\

Ignoring the term $\alpha_1$, $\alpha_2$ and the strata dummies $\omega_j$ which are included to improve precision, (\ref{BLP}) resembles the standard linear regression approach in estimating conditional average treatment effects (\ref{linCATE}). In particular, the coefficient on the (orthogonalized) treatment indicator ($\beta_1$) is the average treatment effect and the coefficient on the interaction term ($\beta_2$) captures heterogeneity in treatment effects. There are, however, three important things that distinguish (\ref{BLP}) from the standard framework.\\

First, the variables of interest are orthogonalized to combat regularization bias in the machine learning proxies.
Subtracting the empirical counterpart of the conditional expectations of $S(Z_i)$ and $D_i$ given $Z_i$ (i.e. $\overline{S(Z)}$ and $p(Z_i)$) from the variables themselves produces orthogonalized  variables. The orthogonalized variables are, by construction, orthogonal to all the covariates in $Z_{i}$, and any function of these, under any $Z_i$-dependent weight. Any confounding effects induced by regularization are therefore partialled out under the weight $w(Z_i)$, as the terms $(D_{i}-p(Z_{i}))(S(Z_{i})-\overline{S(Z)})$ and $(D_{i}-p(Z_{i}))$ are orthogonal to each other and all other functions of $Z_{i}$.\footnote{The orthogonalization technique is closely related to Double/Debiased Machine Learning  \citet{DDML}, and builds on ideas that go all the way back to \citet{FW} and \citet{lovell}. It is clear that orthogonalization is only as good as the covariates available as the orthogonalized variables need not be orthogonal to \textit{unobserved} confounders.}\\

Next, the \textsc{wls} regression is estimated using the main sample ($Data_M$) while the proxies $B(Z_{i})$ and $S(Z_{i})$ are estimated using the auxiliary sample ($Data_A$). If instead the machine learning algorithms used to create the proxies were fit using the entire sample it would be likely to induce a relation between these and the model errors in (\ref{BLP}), where they are used as plug-ins, as data for observation $i$ is used in forming both. This may then lead to a bias in, and slower convergence of, the estimator of $\beta_2$ (\cite{CDDF}). Using different observations to estimate the machine learning
proxies and the regression model breaks any link between the proxies ($B(Z_i)$, $S(Z_i)$) and the model errors ($\epsilon_{i}$). \\


Finally, the interaction term is not limited to a single covariate, and thereby focusing on heterogeneity in treatment effects in a single dimension at a time, which tend to be the case in a standard linear regression framework.
Instead, the (orthogonalized) treatment indicator is interacted with the (orthogonalized) proxy of the \textsc{cate}  which is a function of the entire covariate space and thereby captures heterogeneity in treatment effects across all dimensions of $Z_i$. This latter fact is important as $\beta_2$ can be used to detect heterogeneity across all dimensions of $Z_i$ at once, rather than searching over the covariate space leading to concerns of multiple hypothesis. \\

\textit{Sorted Group Average Treatment Effects}\\
The second feature of $s_0(Z)$ is the Group Average Treatment Effects (\textsc{gates}) which measures the average treatment effects across groups which are defined using $S(Z)$.\footnote{A drawback with using $S(Z)$ to define the groups is that $S(Z)$ is likely to be a biased estimator of the true treatment effects, so the groups defined may not capture the true groupings. The challenge with estimating treatment parameters of this nature is that they depend on the \textit{joint} distribution of the potential outcomes of treated and untreated, which is not identified by randomisation  \citet{altgroup}.}
In particular, we estimate the average treatment effects for four non-overlapping groups $G_1$ to $G_4$ where $G_4$ is the group that consists of the top 25 pct. with the highest predicted treatment effects (most affected) and $G_1$ which consists of individuals with the bottom 25 pct. lowest predicted treatment effects (least affected) according to $S(Z)$. The parameters of interest are then,
\begin{align}
    \mathbb{E}[s_0(Z)|G_k], \quad k=\{1,2,3,4\}
\end{align}
where $G_k$ denotes the $k^{th}$ group. These parameters can be estimated by using \textsc{wls} on the following regression equation,
\begin{align}
    Y_{ij} = \alpha_{0}+\omega_j+\alpha_{1}B(Z_i)+\alpha_{2}S(Z_i)+\sum_{k=1}^4\gamma_{k}(D_i-p(Z_i))1\{i \in G_k\}+\nu_{ij} \label{GATES}
\end{align}
The weights are defined as for (\ref{BLP}) and $1\{i \in G_k\}$ indicates whether individual $i$ belongs to group $k$. Equation (\ref{GATES}) is estimated for both monthly profits and monthly consumption for each RCT and includes strata dummies and uses clustered/robust standard errors as described for equation (\ref{BLP}). The parameters of interest are $\{\gamma_k\}^4_{k=1}$. Importantly, and as described, the treatment indicator is orthogonalized to combat regularization bias, and the regression is estimated using only observations from the main sample ($Data_M$) to avoid overfitting.\\

\textit{Classification Analysis}\\
If the \textsc{blp} and \textsc{gates} detect significant heterogeneity in treatment effects then the Classification Analysis (\textsc{clan}) estimator can be used to describe the average characteristics of the most $(G_4)$ and least $(G_1)$ affected individuals. Information on how effects vary with covariates is useful if the goal is to explore mechanisms or predict what the result of an intervention may be in a specific population. Let $g(Y,Z)$ be a vector of characteristics of an individual. Then the parameters of interest are,
\begin{align}
    \delta_1 = \mathbb{E}[g(Y,Z)|G_1] \quad \text{and} \quad  \delta_4 = \mathbb{E}[g(Y,Z)|G_4]
\end{align}
which measure the average characteristics of the most and least affected individuals. The CLAN estimates are obtained by regressing each characteristic onto a pair of dummies indicating whether an individual belongs to the most or least affected as defined by $S(Z)$,
\begin{align}
    z_{ij} = \delta_{1}1\{i \in G_1\}+\delta_{4}1\{i\in G_4\}+\eta_{ij} \label{CLAN}
\end{align}
where $z_{ij}$ denotes the $j^{th}$ variable in $Z$ that belongs to individual $i\in Data_M$. The CLAN is conducted for both profits and consumption for each RCT using the covariates in Table \ref{Morocco}-\ref{Bosnia}.

\subsubsection{Inference}
To make valid inference on the features of the \textsc{cate} function, it is important to note that the estimators of these, such as $\hat{\beta}_2$, inherits randomness from both the standard \textit{estimation uncertainty} and from the \textit{random splitting} of the sample into a main and auxiliary sample. In particular, the parameters depend implicitly on the auxiliary sample as it is used to estimate the machine learning proxies. Different partitions of the data therefore yield different estimators. \citet{CDDF} show that the estimators are normally distributed conditional on the sample split and under mild regularity conditions. For instance,
\begin{align*}
    \hat{\beta}_{2,s}|Data_A \overset{a}{\sim} N(\beta_{2,s},\sigma_{2,s}^2)
\end{align*}
for $n\in Data_M \xrightarrow{} \infty$. Here the subscript $s$ indicates the split used to estimate $\hat{\beta}_2$. The split-dependent estimator $\hat{\beta}_{2,s}$, along with related confidence intervals, are therefore treated as random \textit{conditional} on the data due to the random partitioning. To address this randomness \citet{CDDF} propose to repeat steps 1-4 and re-estimate equations (\ref{BLP}), (\ref{GATES}) and (\ref{CLAN}) multiple $(S)$ times. The final point estimators and confidence intervals then consists of taking the median of the $S$ split-dependent quantities.\footnote{The median is used, in contrast to using the mean, as the former is more robust to outliers.} In particular, the final point estimator of $\hat{\beta}_2$ is the median of the split-dependent estimators $\hat{\beta}_2=\text{Med}\big{(}\{\hat{\beta}_{2,s}\}^S_{s=1}\big{)}$. Likewise, medians of split-dependent upper and lower confidence bounds define the final confidence interval. Importantly, the nominal confidence bounds have to be adjusted in order to account for both estimation uncertainty and splitting uncertainty induced by random partitioning of the data. Therefore, to guarantee uniformly validity, the final confidence level is $(1-2\alpha)$\%. Similarly, the final sample-splitting adjusted $p$-value is twice the median of the split-dependent $p$-values. \\

The benefit of using multiple sample splits is two-fold. Using many splits is a way to obtain more versions of the estimator of a particular key feature. Letting the final estimator be a function of these estimators (such as the median) helps in utilizing all the data available to estimate the parameters of interest, compared to an estimator obtained from using a single split which utilizes only half of the sample  (\cite{DDML,AtheyTalk}). Another reason for using many splits of the data into main and auxiliary samples is to produce robust estimators. Only using a single split could lead to results that do not hold in general (i.e. for other splits).

\subsection{Elastic Net Regression and the Random Forest Algorithm}
An appealing feature of the method described is that it is generic with respect to the machine learning algorithm used to estimate the proxies $S(Z)$ and $B(Z)$ as it does not rely on strong and specific assumptions on the method employed. To take advantage of this generality we apply two techniques that differ substantially in their approach to predict $s_0(Z)$ and $b_0(Z)$. These are the Elastic Net regularization and the Random Forest algorithm. \\

Elastic Net Regularization is a parametric machine learning algorithm that was developed by \citet{EN} and builds on top of the linear regression framework by regularizing the parameters. In particular, it extends the standard least squares estimator by adding two additional terms to the minimization criterion,
\begin{align}
    \hat{\theta}_{EN}=(1+\lambda_2)\underset{\theta}{\text{argmin}}\{L(\lambda_1,\lambda_2,\theta)\} \label{2.17}
\end{align}
where
\begin{align}
    L(\lambda_1,\lambda_2,\theta)=\|Y-Z\theta\|_2^2+\lambda_2\|\theta\|^2_2+\lambda_1\|\theta\|_1 , \quad \lambda_1,\lambda_2 \geq 0 \label{2.18}
\end{align}
Here, $\theta$ is a $p$-dimensional vector of parameters and $\lambda_1$, $\lambda_2$ are positive constants associated to the L1-norm ($\|\theta\|_1=\sum^p_{j=1}|\theta_j|$) and the L2-norm ($\|\theta\|^2_2=\sum^p_{j=1}\theta_j^2$).
Elastic Net regularization therefore corresponds to a penalized least squares method where the number and the absolute size of the parameters are reduced by the regularization terms $\lambda_1$ and $\lambda_2$. In other words, the method adds automatic variable selection, in which it can select groups of correlated variables, and continuous shrinkage to the standard \textsc{ols} case. Whereas \textsc{ols} will estimate non-zero coefficients $\theta_j$ for all covariates and thereby increasing the chance of overfitting (i.e. fitting noise in the response variables), the Elastic Net will set some coefficients to zero and shrink others to avoid this. The benefit of doing so is an increase in the accuracy of prediction as measured by the mean squared error (\textsc{mse}) as well as a more parsimonious model that increases interpretability  (\cite{EN}). Notably, the Elastic Net works well in high-dimensional settings as it yields a sparse solution vector with only the parameters of the most relevant variables or groups of variables being nonzero, thereby providing an effective data-driven way to detect the variables that appear to be the most important in predicting outcomes linearly. The Elastic Net is better suited to provide an estimator for $s_0(Z)$ if it is well-approximated by a linear combination of pre-specified functions of $Z$. If $s_0(Z)$ is believed to be non-linear other machine learning algorithms are better. \\

A Random Forest is a non-parametric tree-based machine learning algorithm developed by \citet{RF} that builds upon his earlier work on the Classification and Regression Tree (\textsc{cart}) algorithm (\cite{CART}). Random Forests are known to handle both non-linear relations and high-dimensional settings well (\cite{elements}). The Random Forest algorithm is an ensemble method that average over a large number of individual regression trees. In particular, each tree in the Random Forest is estimated on a randomly drawn bootstrap sample, and is composed of partitions of the data based on binary indications of the covariates. These partitions are found using Recursive Binary Splitting (\textsc{rbs}). Each tree then fits a simple model, such as the average outcome, in each of its partitions, and the Random Forest averages over all the tress. Importantly, each tree in the Random Forest uses a random subset of the available covariates $Z$ as potential splitting candidates at each split-point. Growing individual trees on random bootstrap samples with a random subset of covariates available at each splitting point helps in creating decorrelated trees, which leads to lower variance of the resulting Random Forest estimator. In particular, for an individual with a vector of covariates $Z_i$ the estimator is,
\begin{align}
    \hat{Y}_i=\frac{1}{B}\sum^B_{b=1}T_b(Z_i)
\end{align}
where $B$ is the number of trees and $T_b(Z_i)$ is the estimated outcome from the $b^{th}$ tree. The Random Forest improves on individual trees as it is less prone to overfitting and yield more stable estimators. In contrast to the Elastic Net the Random Forest is bad at modeling linearly additive structures (\cite{elements}).

\subsubsection{Evaluation of the Machine Learning Algorithms}\label{bestML}
\citet{CDDF} provide two measures for deciding on the better performing machine learning algorithm in a disciplined way. The best machine learning method for fitting the \textsc{blp} of $s_0(Z)$ is the one that maximizes,
\begin{align}
    \Lambda \equiv |\beta_2|^2\text{Var}(S(Z)) \label{lambda_B}
\end{align}
using the main sample.
Maximizing $\Lambda$ is equivalent to maximizing the correlation between the machine learning proxy predictor $S(Z)$ and the true \textsc{cate} $s_0(Z)$. For \textsc{gates}, the best machine learning algorithm is the one that, given four groups of equal size, maximizes,
\begin{align}
    \bar{\Lambda}\equiv\frac{1}{4}\sum^{K}_{k=1}\gamma_k^2, \quad k\in\{1,2,3,4\} \label{lambda_G}
\end{align}
which captures the part of the variation of $s_0(Z)$ explained by $\widetilde{S(Z)}=\sum^4_{k=1}\gamma_k 1(S(Z)\in G_k)$. This corresponds to maximizing the $R^2$ in a regression of $s_0(Z)$ on $\widetilde{S(Z)}$ without a constant.\\

The measures in (2.20) and (2.21) can only be used for ordinal comparisons as the measures do not tell how well a given algorithm fits the \textsc{blp} or \textsc{gates}, nor how much better a given algorithm is compared to another.
This issue arises as machine learning methods are commonly being evaluated using a \textsc{mse} criteria. As the true treatment effect is never observed such a criterion is infeasible.\footnote{As an exception, \citet{cf} have developed a tree based estimator that uses an adjusted MSE which is an unbiased estimator of the MSE.}

\section{Results}
The identification of heterogeneity in treatment effects from increased access to microcredit, and in particular describing the characteristics of  subgroups of winners and losers, is an important part of understanding the impacts of microcredit and its usefulness as a development tool. This knowledge can be of use in predicting the impacts of access to microcredit in future settings, and thereby help inform policy makers whether or not to allocate resources towards microcredit.\\

By applying the methods described in the previous section we start out by presenting an analysis of the existence of heterogeneous treatment effects on profits and consumption from increased access to microcredit. We then investigate the existence of groups of winners and losers. Finally, given the existence of winners and losers, we depict key characteristics of these subgroups in an attempt to uncover the most important characteristics in predicting heterogeneous effects from increased access to microcredit.\\

Before turning to the actual analysis, Table \ref{MLmethods} summarises the performance of the Elastic Net and Random Forest algorithms. From Table 8 it is evident that the Elastic net outperforms the Random Forest across profits and consumption for both the metrics discussed in section 3.4.1. One possible explanation is that the \textsc{cate} function is linear, or close to linear, and therefore better approximated by the Elastic Net than the Random Forest. An alternative explanation 
is the fact that the Elastic Net is a parametric method opposed to the non-parametric Random Forest. In general, parametric methods require less data to produce precise estimates of the mapping functions from $Z$ to $B(Z)$ and $S(Z)$ compared to non-parametric methods  (see \citet{parametric}). Given the sample sizes considered in this paper this may be the deciding factor. Based on the results of Table \ref{MLmethods}, we choose to focus on the estimates provided by the Elastic Net in the coming subsections. The estimates where the proxies $S(Z)$ and $B(Z)$ are estimated using a Random Forest will be considered as a measure of robustness at the end of the section.\footnote{We implement the Elastic Net (\textsc{en}) and Random Forest (\textsc{rf}) using the \textsc{caret} package in \textit{R}  \citet{caret}. All reported results are medians over 50 sample splits. The tuning parameters for the \textsc{en} are found using twice repeated 2-fold cross-validation and a random grid search. For the \textsc{rf} the number of splitting variables is set to $p/3$ where $p$ denotes the dimension of $Z$, and the number of trees is set to 1000. The outcomes and covariates are rescaled to be between 0 and 1 before estimating the models. The machine learning configurations and the number of sample splits are chosen to accommodate computational time. It takes approx. 18 hours to estimate all quantities at the current settings.}


\subsection{Heterogeneous Effects from Microcredit}
Table \ref{ATEHET} presents estimates of $\hat{\beta}_1$ and $\hat{\beta}_2$. The former measures the average treatment effect (\textsc{ate}) of increased access to microcredit on monthly profits and total monthly consumption both measured in \textsc{usd (ppp)}, and the latter measures the relevance of the proxy $S(Z)$ \textit{and} whether there is heterogeneity in the treatment effects (\textsc{het}). The estimates are obtained from the \textsc{blp} of the \textsc{cate} using equation (\ref{BLP}). Notably, and in line with the original studies, the average treatment effects on profits and consumption across all \textsc{rct}s are not significantly different from zero. Solely focusing on the average effects of increased access to microcredit therefore indicates that it has no effect on the abilities of microentrepreneurs to generate profits. Nor does it impact the well-being of individuals by increasing consumption. \\



Turning to the estimates of $\beta_2$ we observe a different story. In particular for profits, $\hat{\beta}_2$ is positive and significantly different from zero with $p$-values of 0.002 and 0.070 for the \textsc{rct}s in Morocco and Mongolia while it is insignificant for Bosnia \& Herzegovina. This implies that heterogeneity in treatment effects are present in two out of the three \textsc{rct}s considered. The absence of average treatment effects for profits may be due to the presence of groups of winners with high treatment effects and groups of losers with low treatment effects that cancel out when averaging over the entire population. \\

Interestingly, $\hat{\beta}_2$ for consumption remains insignificant across all three \textsc{rct}s, leaving no sign of heterogeneous impacts from increased access to microcredit on total monthly consumption.\\ 

\begin{center}
\captionof{table}{Coefficients of the Best Linear Predictor (BLP) of the CATE across RCTs}\label{ATEHET}
\hspace{0 cm}
\hbox{
\scalebox{0.75}{
\begin{tabularx} {1.33\textwidth}{lcc|cc|cc} \hline \hline \\
 & \multicolumn{2}{c}{\textbf{Morocco}} & \multicolumn{2}{c}{\textbf{Mongolia}} & \multicolumn{2}{c}{\textbf{Bosnia \& Herzegovina}} \\
 & ATE ($\hat{\beta}_1$) & HET ($\hat{\beta}_2$) & ATE ($\hat{\beta}_1$) & HET ($\hat{\beta}_2$) & ATE ($\hat{\beta}_1$) & HET ($\hat{\beta}_2$) \\ \cmidrule{2-7}
Profits &26.959 & 0.260 & -1.893 & 0.186 & 74.259 & 0.088
\\
& (-25.386,77.802) & (0.104,0.423) & (-4.017,0.277) & (0.012,0.357) & (-62.844,213.014) & (-0.317,0.518)
 \\
 & [0.607] & [0.002] & [0.180] & [0.070] & [0.497] & [0.930]\\
Consumption & -14.078 & 0.146
 & 109.152 & 0.279 & -38.654 & 0.069 \\
& (-36.867,8.542) & (-0.066,0.356) & (-46.484,255.036) & (-0.107,0.699) & (-183.837,113.039) & (-0.222,0.409) \\
& [0.424] & [0.383] & [0.343] & [0.271] & [1.000] & [0.946] \\
& \multicolumn{6}{c}{} \\  \hline \hline
\end{tabularx}}}
\begin{tablenotes}[para,flushleft]
\scriptsize
 Notes$:$
 The table presents estimates of the average treatment effects (\textsc{ate}) and heterogeneity in treatment effects (\textsc{het}) for profits and consumption. The estimates are given by the median of the estimates of $\beta_1$ and $\beta_2$ obtained by applying \textsc{wls} to equation (\ref{BLP}) over 50 splits. Similarly, the sample splitting adjusted confidence intervals and $p$-values are obtained by the median over 50 splits with the $p$-values being twice this value to correct for sample splitting uncertainty. Clustered standard errors at the village level for Morocco and Mongolia, and robust standard errors for Bosnia \& Herzegovina are used. Strata dummies at the village level for Morocco and at the province level for Mongolia are included. The proxies S(Z) and B(Z) that enters equation (\ref{BLP}) are estimated using an Elastic Net on the covariates presented in Tables \ref{Morocco}, \ref{Mongolia}, and \ref{Bosnia}, for Morocco, Mongolia and Bosnia \& Herzegovina respectively. Profits and consumption are monthly profits from business activities and total monthly consumption, both measured in \textsc{usd ppp}  indexed to 2009. \newline
 $\bullet$ 90\% confidence intervals in parentheses.\newline
 $\bullet$ $p$-values for the hypothesis that the parameter is equal to zero in brackets.\newline
\end{tablenotes}
\end{center}

\subsection{Groups of Winners and Losers}
\begin{figure}
\centering

\scalebox{0.9}{
\begin{subfigure}[b]{1\textwidth}
   \centering
          \caption{Morocco}
   \includegraphics[width=1\linewidth]{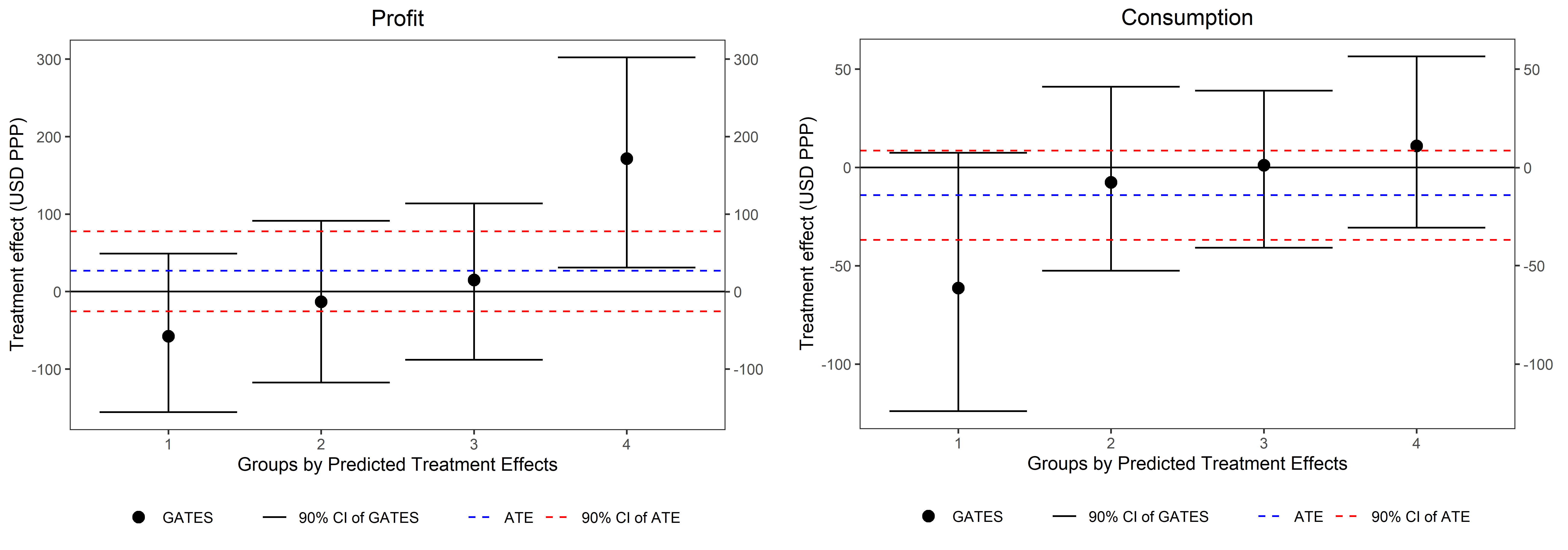}
   \label{fig:MOR}
\end{subfigure}}

\scalebox{0.9}{
\begin{subfigure}[b]{1\textwidth}
    \caption{Mongolia}
   \includegraphics[width=1\linewidth]{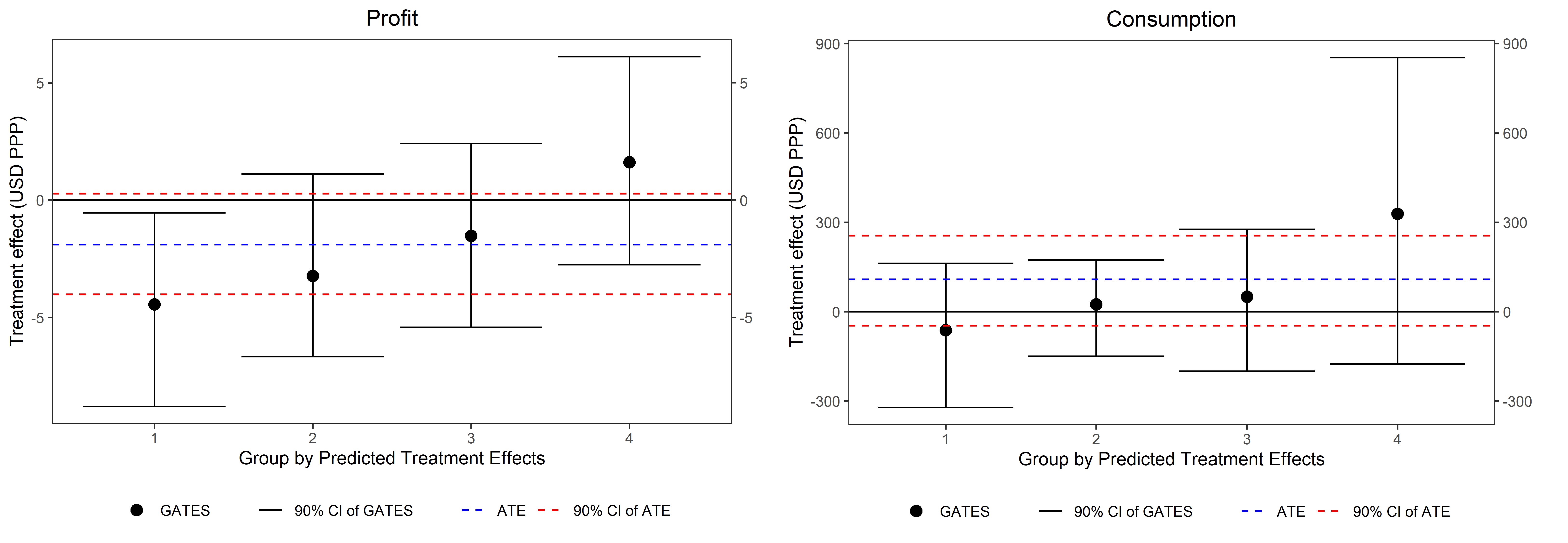}
   \label{fig:MON}
\end{subfigure}}

\scalebox{0.9}{
\begin{subfigure}[b]{1\textwidth}
   \begin{center}
          \caption{Bosnia \& Herzegovina}
   \end{center}
   \includegraphics[width=1\linewidth]{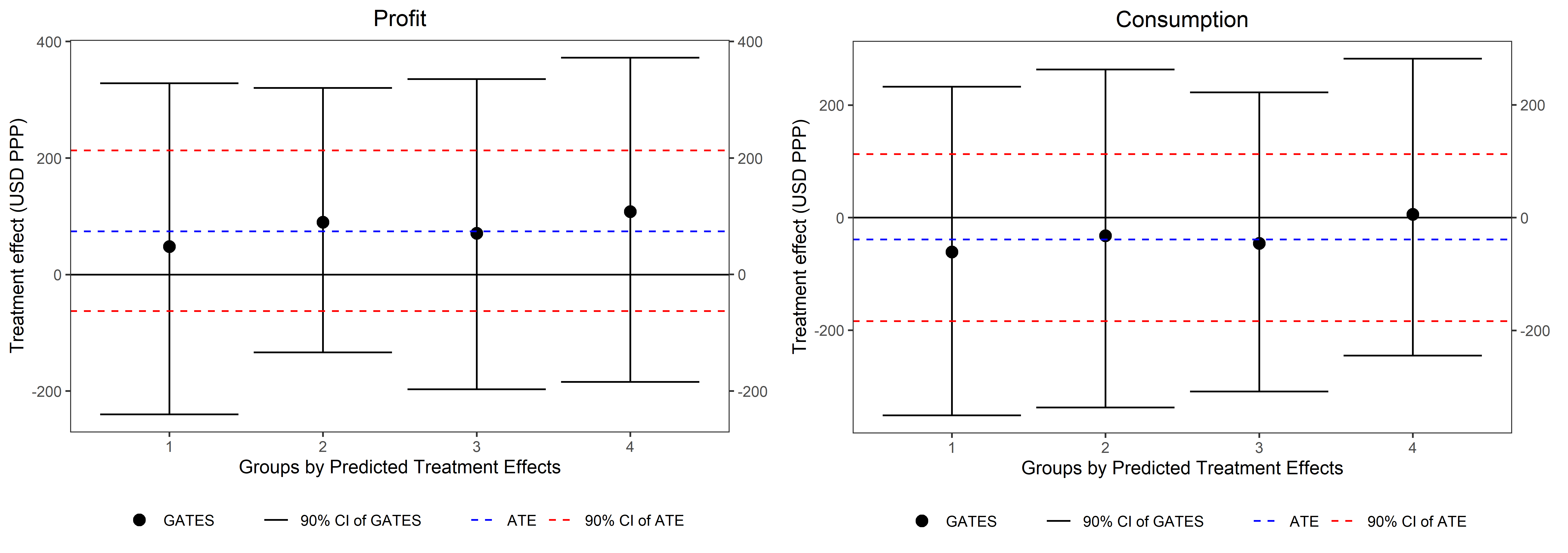}
   \label{fig:BOS}
\end{subfigure}}

\caption[]{Sorted Group Average Treatment Effects (GATES) for profits and consumption across RCTs}\label{GATESfig}
\begin{tablenotes}[para,flushleft]
\scriptsize
 Notes$:$ Figures (a), (b) and (c) presents the Sorted Group Average Treatment Effects (GATES) point estimates for monthly profits and monthly consumption measured in \textsc{usd (ppp)} indexed to 2009 along with 90 pct. confidence intervals for the RCTs in Morocco, Mongolia and Bosnia \& Herzegovina, respectively. The final estimates and confidence intervals are found by found by estimating (\ref{GATES}) using WLS over 50 splits and then taking the median of the split-dependent estimates. Clustered standard errors at the village level for Morocco and Mongolia, and robust standard errors for Bosnia \& Herzegovina are used. Strata dummies at the village level for Morocco and at the province level for Mongolia are included. The four non-overlapping groups G1 to G4 are defined by the proxy $S(Z)$ which is estimated using an Elastic Net. G4 consists of the top 25 pct. most affected and G1 of the bottom 25 pct. least affected. The figures also show the Average Treatment Effect (ATE) along with 90 pct. confidence intervals.\newline
\end{tablenotes}
\end{figure}

To analyse the nature of the heterogeneity (or lack thereof) Figure \ref{GATESfig} shows the point estimates and related 90 pct. confidence intervals for the Sorted Group Average Treatment Effects (\textsc{gates}) across four groups G1 to G4: G4 consists of top 25 pct. of the individuals with the highest treatment effect and G1 the bottom 25 pct. all defined by the proxy $S(Z)$.\footnote{As mentioned in section \ref{features}, G1-G4 may not reflect the true groupings when defined using $S(Z)$. The positive estimates of $\hat{\beta}_2$ and the monotonically increase in group treatment effects depicted in Figure \ref{GATESfig} indicate that $S(Z)$ and $s_0(Z)$ are positively correlated. Moreover, there is evidence that $S(Z)$ do capture groups that differ substantially in their treatment effects, which is the main object of interest.} In particular, it is evident that there are no heterogeneity in treatment effects on consumption as there is substantial overlap 
between the confidence intervals of the treatments effects for all groups across all \textsc{rct}s. The idea that microcredit may benefit poor people by allowing them to raise their consumption and thereby their well-being therefore seems to be unobtainable, at least in the short run, when considering both distributional and average impacts given the three \textsc{rct}s examined in this paper. \footnote{Given that our focus is upon total monthly consumption we cannot rule out that individuals shift consumption spending across goods.}\\

\vspace{0.1cm}For profits the \textsc{gates} estimates in Figure \ref{GATESfig} do indicate that heterogeneity in treatments exist. For the \textsc{rct} in Morocco it comes about as a group of winners whereas there are significant losers for the \textsc{rct} in Mongolia, thereby supporting both the negative and positive claims on the effects of microcredit. Finally, there is no evidence of heterogenous impacts for the \textsc{rct} in Bosnia \& Herzegovina. It is difficult to  explain  this finding given that the \textsc{rct}s differ across a variety of parameters (see Table \ref{t1}). One obvious difference, however, is that the \textsc{rct} in Bosnia \& Herzegovina is 
randomised at the individual  level by focusing on marginal clients all over Bosnia \& Herzegovina. This contrasts the \textsc{rct}s in Morocco and Mongolia that randomised at the community level and thereby created strata of treated and untreated people. Any spill-overs, general equilibrium effects, or social network effects inside villages are therefore captured by only the latter two.  \citet{india} argue that spill-overs are important, while \citet{Morocco} argue for it having limited effects. Nevertheless, if spill-overs are important drivers of treatment effects it won't be captured by the sampling and randomisation schemes used in Bosnia \& Herzegovina.

\newgeometry{top=0.5cm, bottom=0.5cm, left=0.5cm, right=0.5cm}
\singlespacing
\onehalfspacing
\restoregeometry
\newpage

Table \ref{Gatestable} shows the estimated treatment effects on profits and consumption for the most and least affected as well as the difference between the groups. Notably, the most affected in Morocco experience a significant increase in profits of 171.7 \textsc{usd (ppp)} or 103 percent of the average value in the control group. Comparing this to the least affected with a point estimate of -57.7 \textsc{usd (ppp)} gives a significant difference of 227 \textsc{usd (ppp)}. In other words, increased access to microcredit seem to generate groups of winners and losers in Morocco. \\


Turning to Mongolia shows that increased access to microcredit can have harmful effects for a part of the population as the least affected individuals experiences a statistically significant drop in monthly profits of 4.4 \textsc{usd (ppp)}, corresponding to a decrease of  102 percent of the average value in the control group. Here, it is worth remembering that the \textsc{rct} in Mongolia focused on the poorest of the poor women (see Table \ref{t1}). Additionally, when comparing the effects of the most and least affected individuals in Mongolia we observe a significant difference just shy of 7 \textsc{usd (ppp)}, indicating that increased access to microcredit produces groups of winners and losers.
 Finally, for Bosnia \& Herzegovina there is no significant difference. \\

The heterogeneity in treatment effects on monthly profits indicated by the estimates of the \textsc{blp} of the \textsc{cate} represents  the presence of winners and losers across the \textsc{rct}s in Morocco and Mongolia. Heterogeneous treatment effects may be driven by gains of entrants at the expense of incumbent business owners. Alternatively, existing business owners may reap most of the benefits from increased access to microcredit due to expertise or skill, whereas unexperienced entrants see most of their profits evaporate in the presence of high interest rates on microloans (see Table \ref{t1}). \\

To gain a better understanding of the heterogeneity in treatment effects we turn to the Classification Analysis (\textsc{clan}) which describes the average characteristics of the most and least affected individuals. If there exist characteristics that are pronounced in either group across the \textsc{rct}s, then knowledge of such characteristics could be of relevance in predicting whether access to microcredit is favourable or not, and in particular who it may help and who it may harm in future studies.


\restoregeometry
\newgeometry{top=3cm, bottom=3cm, left=2cm, right=2cm}
\begin{center}
\captionof{table}{GATES estimates for profits and consumption across RCTs}\label{Gatestable}
\hspace{0.5 cm}
\hbox{
\scalebox{0.9}{
\begin{tabularx} {0.9\textwidth}{lccc} \hline \hline
& \multicolumn{3}{c}{\textbf{Panel A: Morocco}} \\
 & 25 \% most ($\hat{\gamma}_{4}$) & 25 \% least ($\hat{\gamma}_1$) & Difference ($\hat{\gamma}_{4}$ - $\hat{\gamma}_{1}$) \\\cmidrule{2-4}
Profits & 171.739
 & -57.724 & 226.847  \\
& (31.203,302.328) & (-155.692,49.026) & (49.537,409.779) \\
& [0.030]& [0.591]& [0.028] \\
Consumption & 10.934 & -61.296 & 76.580  \\
& (-30.591,56.462) & (-124.001,7.433) & (-8.028,158.341) \\
& [0.853] & [0.162] & [0.156] \\\cmidrule{2-4}
& \multicolumn{3}{c}{\textbf{Panel B: Mongolia}} \\
 & 25 \% most ($\hat{\gamma}_{4}$) & 25 \% least ($\hat{\gamma}_1$) & Difference ($\hat{\gamma}_{4}$ - $\hat{\gamma}_{1}$) \\\cmidrule{2-4}
Profits & 1.609 & -4.442 & 6.561 \\
& (-2.746,6.115) & (-8.798,-0.539) &  (1.387,12.048) \\
& [0.840] & [0.049] & [0.039]
 \\
Consumption & 328.517 & -61.879 & 392.856\\
& (-174.687,853.155) & (-321.696,162.486) & (-281.516,1051.871) \\
& [0.371] & [1.000] & [0.554] \\\cmidrule{2-4}
& \multicolumn{3}{c}{\textbf{Panel C: Bosnia}} \\
 & 25 \% most ($\hat{\gamma}_{4}$) & 25 \% least ($\hat{\gamma}_1$) & Difference ($\hat{\gamma}_{4}$ - $\hat{\gamma}_{1}$) \\\cmidrule{2-4}
Profits & 107.882 & 47.919 & 50.830 \\
& (-184.485,372.267) & (-240.218,328.258) &  (-365.110,431.934) \\
& [0.824] & [1.000] & [1.000] \\
Consumption & 5.851 & -60.888 & 66.265\\
& (-244.900,282.417) & (-351.425,232.547) & (-358.288,459.754) \\
& [1.000] & [1.000] & [1.000]\\ \hline \hline
\end{tabularx}}}
\begin{tablenotes}[para]
\scriptsize	
Notes$:$ The table presents estimates of the Sorted Group Average Treatment Effects (\textsc{gates}) for monthly profits and monthly consumption measured in \textsc{usd (ppp)} indexed to 2009. Panel A, B and C shows the estimates for the \textsc{rct}s in Morocco, Mongolia and Bosnia \& Herzegovina respectively. The final estimates, confidence intervals and $p$-values are found by estimating (\ref{GATES}) by \textsc{wls} over 50 splits and then taking the median of the split-dependent estimates. Clustered standard errors at the village level for Morocco and Mongolia, and robust standard errors for Bosnia \& Herzegovina are used. Strata dummies at the village level for Morocco and at the province level for Mongolia are included. The two non-overlapping groups G1 and G4 are defined by the proxy $S(Z)$ which is estimated using an Elastic Net. G4 consists of the 25 pct. most affected and G1 of the 25 pct. least  affected. The difference is found by testing whether the group estimates are equal. The final estimate, confidence interval and $p$-value for this quantity are found in the same way as the estimates of $\gamma_1$ and $\gamma_4$. \newline
 $\bullet$ 90\% confidence intervals in parentheses.\newline
 $\bullet$ $p$-values for the hypothesis that the parameter is equal to zero in brackets.\newline
\end{tablenotes}
\onehalfspacing
\end{center}

\newpage

\subsubsection*{Household Characteristics or Aggregate Factors}
Before turning to the \textsc{clan} estimates it is of interest to analyse the relative role of sociodemographic factors in explaining the heterogeneity discovered so far. To do so, we analyse whether the heterogeneity is primarily explained by household level characteristics, or by aggregate or village level factors. The villages considered in both \textsc{rct}s differ on a number of parameters including location, population density, infrastructure, access to education, and many more. For the \textsc{rct} in Mongolia a few village characteristics are available (see Table \ref{Mongolia}). However, for the \textsc{rct} in Morocco no village characteristics are available. To overcome this, we use village/province-level strata dummies as proxy variables for aggregate-level characteristics. These dummies also capture other   factors such as the efficacy of the \textsc{mfi}s branch managers.\\

To quantify the relative importance of the household and aggregate level covariates in predicting heterogeneity in treatment effects we report the adjusted $R^2$ from regressing a dummy  that equals 1 if an individual belongs to the most affected and 0 if an individual belongs to the least affected, defined from the quartiles of the proxy $S(Z)$, on aggregate level covariates (including strata dummies), household level covariates, and all covariates, respectively.
From Table \ref{HHvsAGG} it is evident that household covariates explain much of the variation in the heterogeneity in both \textsc{rct}s as household level covariates explain 81 (49) percent of the variation in Mongolia (Morocco). In comparison, aggregate level factors such as manager quality, spillover effects and general equilibrium effects explain 67 and 87 percent of the variation in Mongolia and Morocco, respectively. Even though it is not clear-cut which factors that are most important, it seems as if household covariates are informative in explaining the heterogeneity in treatment effects in both \textsc{rct}s, and hence the forthcoming \textsc{clan} analysis is indeed of relevance.

\subsection{Group Characteristics}
As there are no sign of heterogeneity in treatment effects on profits in Bosnia \& Herzegovina, nor for consumption across all three \textsc{rct}s we limit the \textsc{clan} to monthly profits in Morocco and Mongolia, and Table \ref{CLANtable} presents the estimates. In particular, we report the \textsc{clan} estimates of the five baseline covariates that have the highest correlation with the proxy $S(Z)$ as these seem to be the most important in \textit{predicting} heterogeneity.\footnote{Due to regularization in the machine learning algorithms used it is not possible to infer causal relations from the covariates. Regularization focus on some covariates and not others based on correlation and it is hard to infer whether the actual choice cause or correlates with the outcome  \citet{AtheyTalk}.} This is done to avoid any concerns about the choice of covariates presented in the \textsc{clan} being arbitrary and thereby the possibility of searching over variables. We choose to limit the \textsc{clan} to five variables for each \textsc{rct} to reduce the number of hypothesis tests. \\

In Panel A and Panel B of Table \ref{CLANtable}, we observe that the most important predictors of heterogeneity relates to self-employment activities, consumption levels, and indebtedness across both \textsc{rct}s. In particular, the most affected across the \textsc{rct}s seem to be less likely to be engaged in loan activities prior to the trials compared to the least affected. In Morocco 1 out of 6 of the most affected had outstanding loans compared to just shy of 1 out of 3 of the least affected, and 1 out of 5 of the people in the control group.  Similarly, in Mongolia, the most affected had a current debt that was 857.5 \textsc{usd (ppp)} lower than the least affected at baseline. This difference may result from the most affected being less likely to borrow (as in Morocco) or because they borrow smaller amounts. In any case, the magnitude of the loan activities of the most affected are significantly lower compared to the least affected in Mongolia. It is important, however, to point out that the most affected are still slightly more indebted than the control group. 
As no \textsc{mfi} were operating in any of the villages in either country prior to the trials the difference in borrowings between the most and least affected indicates that individuals belonging to the former group were less likely to engage in loan activities other than microcredit. Such other activities are primarily made up of informal loans from local money-lenders or relatives but other sources are also recorded such as outstanding loans from utility companies. \\


Across the \textsc{rct}s a number of self-employment activity related variables are important predictors of the heterogeneity in treatment effects on profits. The fact that previous self-employment activities are related to the treatment effects is consistent with the findings of \citet{meagerUNDER}, previous business owners seem to gain the most from access to microcredit across \textsc{rct}s, and \citet{hettreat} who finds that long-term effects on business scale and performance also depend on previous business ownership. The \textsc{clan}, however, allows for a more detailed description of the factors related to self-employment that seem to be important in predicting heterogeneity. In particular, the most affected in Morocco are business owners that are more likely to be women (8.3 pp.), less likely to be engaged in animal husbandry (18.4 pp.) and more likely to be running a business that is not related to agriculture (4.9 pp.) compared to the least affected.\footnote{These findings also hold , although to a lesser extent, when compared to the control group.} In the case of Mongolia although the most affected are more likely to be women with prior business experience (16.9 pp.), they are also more likely to receive income from agricultural activities (12.3 pp.).\footnote{Recall that the study in Mongolia focus on women. Households with prior business experience are therefore necessarily female. It is, however, uncertain with the data at hand whether women would outperform men.}   Moreover, the most affected had significantly smaller businesses as measured by monthly business revenues at baseline as the revenues are 634 \textsc{usd (ppp)} which is almost 730 \textsc{usd (ppp)}  lower than those of the least affected, and 143 \textsc{usd (ppp)} lower than the control group.  This latter finding could indicate that it is small business owners that are able to expand their businesses when given access to microcredit that gain the most in Mongolia. \\


In summary, it seems as if previous self-employment activities are important in predicting winners and losers across \textsc{rct}s. Apart from women with previous business experience, the exact nature of the self-employment activities are, on the other hand, harder to generalize across the \textsc{rct}s. If previous self-employment activities work as moderators for the causal effect of increased access to microcredit then it is not surprising to find differences in the importance of the specific self-employment activities as these may operate with different strengths and effectiveness across contexts (\cite{Deaton1}). It may also be the case that some common characteristics are missed when considering only the five most effective predictors of heterogeneity.\\

Finally, it seems as if the most affected across both 
\textsc{rct}s have lower monthly consumption compared to the least affected. In Morocco, the most affected consume 371 \textsc{usd (ppp)} on average each month which is 123 \textsc{usd (ppp)} lower than the least affected, and slightly lower than the control group (see Table \ref{Morocco}). In Mongolia, the point estimate of the difference between most and least affected is -44 \textsc{usd (ppp)} but this is insignificant at conventional significance levels. Total monthly consumption therefore seems to be an important predictor of heterogeneity, and there are weak indications towards the most affected having lower consumption levels across settings. Lower consumption, if driven by lower consumption of temptation goods, could imply that households are more patient or disciplined in their spending. Indeed, the lower consumption of the most affected in Morocco is driven by significantly lower consumption of non-durables (not shown in Table \ref{CLANtable}). However, this finding does not apply for the most affected households in Mongolia.

\newpage

\begin{center}
\captionof{table}{Classification Analysis (CLAN) for profits across RCTs}\label{CLANtable}
\hspace{0 cm}
\hbox{
\scalebox{0.75}{
\begin{tabularx} {1.33\textwidth}{lccc} \hline \hline
& & & \\
& \multicolumn{3}{c}{\textbf{Panel A: Morocco}} \\
 & 25 \% most ($\hat{\delta}_{4}$) & 25 \% least ($\hat{\delta}_1$) & Difference ($\hat{\delta}_{4}$ - $\hat{\delta}_{1}$) \\\cmidrule{2-4}
Total monthly consumption of household (USD) & 371.250 & 490.974 & -123.132 \\
& (343.631,398.706) & (463.194,518.755) & (-162.309,-84.504) \\
Self-emp. activity run by women & 0.174 & 0.089 & 0.083 \\
& (0.148,0.200) & (0.063,0.114) & (0.047,0.120) \\
Household has animal husbandry as self-emp. activity & 0.361 & 0.561 & -0.184  \\
& (0.323,0.398) & (0.524,0.598) & (-0.237,-0.131)  \\
Household has non-agriculture as self-emp. activity & 0.172 & 0.129 & 0.049  \\
& (0.145,0.198) & (0.101,0.158) & (0.010,0.088) \\
Household has any outstanding loans & 0.167 & 0.294 & -0.130  \\
& (0.135,0.200) & (0.262,0.326) & (-0.174,-0.087)
 \\\cmidrule{2-4}
& \multicolumn{3}{c}{\textbf{Panel B: Mongolia}} \\
 & 25 \% most ($\hat{\delta}_{4}$) & 25 \% least ($\hat{\delta}_1$) & Difference ($\hat{\delta}_{4}$ - $\hat{\delta}_{1}$) \\\cmidrule{2-4}
Current debt of household (USD) & 519.697 & 1393.255 & -857.551
 \\
& (152.503,873.556) & (1039.424,1738.899)  & (-1338.041,-360.266) \\
Respondent receives income from agricultural activities & 0.178 & 0.059 & 0.123 \\
& (0.123,0.233) & (0.006,0.112) & (0.037,0.200)
\\
Household has a business activity & 0.644 & 0.492 & 0.169 \\
& (0.556,0.732) & (0.404,0.580) & (0.044,0.294)   \\
Total monthly revenues of respondents business (USD) & 634.100 & 1307.063 & -729.946 \\
& (161.356,1094.912) & (870.046,1748.833) & (-1393.561,-77.445)   \\
Total monthly consumption of household (USD) &  564.639 & 614.079 & -44.380
 \\
& (462.490,751.718) & (501.044,765.852)  & (-182.971,87.231) \\
& & & \\
\hline \hline
\end{tabularx}}}
\begin{tablenotes}[para,flushleft]	
\scriptsize	
Notes$:$ The \textsc{rct} estimates are found by estimating equation (\ref{CLAN}) over 50 splits and taking the median of the estimates. In particular, the estimates for the most and least affected are found by regressing the given (baseline) covariate on two indicators using the main samples. The first equals 1 if the individual belongs to the top 25 pct. who experienced the largest treatment effect and the second equals 1 if an individual belong to the bottom 25 pct who experienced the lowest treatment effect as measured by the Elastic Net Proxy $S(Z)$. The estimate of the difference in treatment effects is found by testing whether the coefficients of the before mentioned indicators are equal to each other and taking the median. All the covariates presented in the table were recorded at baseline. Panel A and B present the estimates for the RCTs Morocco and Mongolia, respectively.
The covariates presented for each panel are the five variables that have the highest correlation with the Elastic Net proxy predictor $S(Z)$. All monetary measures are in \textsc{usd (ppp)} indexed to 2009.
\newline
 $\bullet$ 90\% confidence intervals in parentheses.\newline
\end{tablenotes}
\onehalfspacing
\end{center}
\newpage


\subsection{Robustness: Alternative Machine Learning Algorithm}
As a measure of robustness we consider the estimates obtained from using a Random Forest to estimate the proxies $B(Z)$ and $S(Z)$. Given that in some cases the Random Forest algorithm  outperformed Elastic Net (see Table \ref{MLmethods}), it is worth analysing whether the estimates are robust to a change of method. Tables \ref{RFappend1} and \ref{RFappend2} present the \textsc{blp} and \textsc{gates} estimates obtained from using proxies $B(Z)$ and $S(Z)$ from a Random Forest. Crucially, any sign of heterogeneity discovered using the Elastic Net disappears when the proxies are estimated using a Random Forest. The $\hat{\beta}_2$ estimates for consumption and profits from (\ref{BLP}) are insignificant across all \textsc{rct}s. Moreover, and in line with this finding, there are no significant difference between the most and least affected for profits and consumption across all \textsc{rct}s. In addition, cases where Random Forest appear to perform better than the Elastic Net give no indication of heterogeneity in treatment effects.In addition, cases where Random Forest appear to perform better than the Elastic Net give no indication of heterogeneity in treatment effects.\\


It is important to underline that in this study our results  depend crucially on the machine learning algorithm used for estimating the proxies $B(Z)$ and $S(Z)$. In this context, the measures to determine the better performing machine learning algorithm developed by \citet{CDDF}, and discussed in Section \ref{bestML}, provide a disciplined way for the analyst to counter the uncertainty as to which algorithm to deploy. It does, however, not rule out that alternative and better performing methods were implemented but disregarded due to null-findings. 
One possibility is to introduce pre-analysis plans that specify the machine learning algorithms to be applied. Therefore, rather than substituting for pre-analysis plans, the method developed by \citet{CDDF} may in fact be a complement to them, as it limits the restrictions of the pre-analysis plan to more general factors, such as a list of the machine learning algorithms to be used. These machine learning algorithms can in turn be chosen to cover each others strengths and weaknesses which is often known in advance to some degree.
\pagebreak

\section{Concluding Remarks}
Microcredit has been both praised and vilified as a development tool. Understanding heterogeneity in the effects of microcredit is therefore key in informing policy makers whether microcredit is worth allocating resources towards as a means to alleviate poverty. This paper has sought to uncover heterogeneous impacts by presenting evidence from three very different \textsc{rct}s using modern machine learning methods.  \\

Across the \textsc{rct}s we find no evidence of distributional impacts on consumption from increased access to microcredit. Coupling this with zero average effects across all settings therefore points towards microcredit having no effect on household well-being through increases or decreases in consumption. On the contrary the absence of average effect on profits from increased access to microcredit do seem to mask the presence of heterogenous effects. Across two of the three \textsc{rct}s we find evidence of heterogeneous effects on profits which present itself as groups of winners and losers thereby supporting both the negative and positive claims of microcredit. We find that the heterogeneity in treatment effects can be predicted from household level covariates and that the most affected households seem to be less indebted, engage in self-employment activities and consume less as baseline. These findings could be of use in analysing what the effects of microcredit may be in future settings in which the targeted population has different characteristics.\\





We observe that the findings of heterogeneous effects are not robust to the choice of machine learning method. This raises the  possibility of researchers searching over algorithms to validate specific preconceived notions. One way to avoid such concerns is to pre-register the analysis, including the machine learning algorithms to be applied in conjunction with the method developed by \citet{CDDF}. This would limit concerns of model dependence while still offering a disciplined way for researchers to leverage analyses of heterogeneous treatment effects through the use of multiple machine learning methods.

\newpage
\newgeometry{left=2.5cm, right=2.5cm,top=3cm, bottom=3cm}
\normalem
\printbibliography
\newpage


\newgeometry{top=2cm, right=2cm ,left=1.7 cm, bottom=0.5cm}

\begin{center}
\singlespacing
\captionof{table}{Baseline covariates used for detecting heterogeneity for the RCT in Morocco}\label{Morocco}
\hspace{0 cm}
\hbox{
\scalebox{0.75}{
\begin{tabularx}{1.28\textwidth}{lccccccc} \hline \hline \\
 &  & \multicolumn{3}{c}{\textbf{Control group}} & & \multicolumn{2}{c}{\textbf{Treatment - Control}}  \\ \cline{3-5} \cline{7-8}
 & Total obs. & Obs. & Mean & SD & & Coefficient & \textit{p}-value \\\hline
\textit{Household Composition} &  &  &  &  &  & &  \\
Members resident & 5329 & 2687 & 3.837 & 3.216 &  & -0.049 & 0.822 \\
Number of adults (age 16 or older)  & 5329 & 2687 & 2.560 &2.261  &  & -0.018& 0.900
  \\
Head age  & 5329 & 2687 & 35.470 &25.008  &  &-0.029 &0.987
  \\
Male head & 5329 & 2687 & 0.697 & 0.460 &  &-0.011 &0.713
  \\
Head with no education  & 5329 & 2687 & 0.463 & 0.499 &  & -0.019& 0.518 \\
   &  &  &  &  &  & &  \\
\textit{Access to Credit} &  &  &  &  &  & & \\
Has borrowed from any source  & 5329 & 2687 & 0.193 & 0.395  &  & 0.025 & 0.508

  \\
Total amount of outstanding loans  & 5329 & 2687 & \$195.055
& \$678.525 &  & \$10.776 & 0.768 \\
   &  &  &  &  &  &  \\
\textit{Self-employment activities} &  &  &  &  &  & & \\
Has self-employment activity  & 5329 & 2687 & 0.578 & 0.494 &  & -0.027 & 0.454 \\
Animal husbandry self-emp. activity  & 5329 & 2687 & 0.390 & 0.488 &  &0.025 & 0.458 \\
Agricultural self-emp. activity  & 5329 & 2687 & 0.438 & 0.496 &  &0.009 & 0.793  \\
Non-Agricultural self-emp. activity & 5329 & 2687 & 0.156 & 0.363 &  &-0.035 & 0.073 \\
Number of activities managed by women  & 5329 & 2687 &0.116  & 0.321 &  & 0.002 & 0.936
  \\
    &  &  &  &  &  &  & \\
 \textit{All businesses} &  &  &  &  &  &  &\\
Total profits from self-emp. (past 12 months)  & 5329 & 2687 & \$1189.550 & \$8282.596 &  & -\$207.538 & 0.560  \\
Total expenses from self-emp. (past 12 months)  & 5329 & 2687 & \$2390.893 & \$7288.378 &  & -\$293.282 & 0.281 \\
Total output from self-emp. (past 12 months)  & 5329 & 2687 & \$3580.443 & \$10491.722&  & -\$500.820 & 0.292 \\
Total current stock of assets related to self-emp.  & 5329 & 2687 & \$2064.711 & \$4121.195 &  & \$98.681 & 0.669  \\
    &  &  &  &  &  &  & \\
 \textit{Farms} &  &  &  &  &  &  &\\
Investment (past 12 months)  & 5329 & 2687 & \$1.908 & \$12.715 &  & \$0.146 & 0.795 \\
Sales (past 12 months)  & 5329 & 2687 & \$1363.546 & \$6468.694 &  & \$79.829 & 0.803  \\
Expenses (past 12 months)  & 5329 & 2687 & \$518.910 & \$1499.275 &  &\$84.112 & 0.331 \\
Savings   & 5329 & 2687 & \$202.106 & \$651.827 &  & \$4.702 & 0.923 \\
Days worked (past 12 months)  & 5329 & 2687 & 44.339 & 90.157 &  & 4.157& 0.535\\
    &  &  &  &  &  &  & \\
 \textit{Animal Husbandry} &  &  &  &  &  &  &\\
Investment (past 12 months)  & 5329 & 2687 & \$61.119 &  \$337.831 &  & \$1.421 & 0.902
 \\
Sales (past 12 months)  & 5329 & 2687 & \$545.986 & \$1647.877 &  & \$33.509& 0.692  \\
Expenses (past 12 months)  & 5329 & 2687 & \$636.300 & \$1980.628 &  & \$78.348& 0.424 \\
Savings   & 5329 & 2687 & \$1700.391 & \$3370.306 &  & \$141.433 & 0.479 \\
Days worked (past 12 months)  & 5329 & 2687 & 81.153 & 142.850 &  & 4.595 & 0.633 \\
    &  &  &  &  &  &  & \\
 \textit{Consumption} &  &  &  &  &  &  &\\
Total monthly consumption  & 5329 & 2687 & \$383.800
 & \$342.228 &  & -\$1.810
 & 0.935 \\
Monthly expenditure on durables  & 5329 & 2687 & \$7.242 & \$43.819 &  & \$0.603 & 0.733 \\
Monthly expenditure on non-durables  & 5329 & 2687 & \$376.558 &\$331.803 &  & -\$2.413& 0.910 \\
Household per capita cons. is among lowest 25 pct.  & 5329 & 2687 & 0.193 & 0.395 &  & -0.007& 0.774
  \\
    &  &  &  &  &  &  & \\
 \textit{Other} &  &  &  &  &  &  &\\
Total income from any source (past 12 months)  & 5329 & 2687 & \$3598.951 & \$9623.042 &  & \$248.820 & 0.562 \\
Household owns land  & 5329 & 2687 & 0.405 & 0.491 &  & 0.010 & 0.771 \\
Household rents land  & 5329 & 2687 & 0.063 & 0.243 &  & -0.003 & 0.783 \\
Distance in km from home to marketplace  & 5329 & 2687 & 8.569 & 9.457 &  & -0.780  & 0.447  \\
   &  &  &  &  &  & &  \\ \hline \hline
\end{tabularx}}}
\begin{tablenotes}[para]
\scriptsize{}	
 Note: All variables are recorded at baseline (pre-treatment) and all monetary variables are measured in \textsc{usd} (\$) \textsc{pp} indexed to 2009. In addition to the household characteristics presented in the table 81 village-pair dummies are included to capture effects at the aggregate level. Following the authors of the original study we derive the coefficient and $p$-value by regressing the respective baseline covariate on the treatment dummy and using standard errors clustered at the village level.\newline
\end{tablenotes}
\onehalfspacing
\end{center}
\restoregeometry
\newpage

\newgeometry{top=0.2cm, right=2cm ,left=1.7 cm, bottom=0.2cm}
\begin{center}
\singlespacing
\captionof{table}{Baseline covariates used for detecting heterogeneity for the RCT in Mongolia}\label{Mongolia}
\hspace{0 cm}
\hbox{
\scalebox{0.75}{
\begin{tabularx}{1.33\textwidth}{lccccccc} \hline \hline \\
 &  & \multicolumn{3}{c}{\textbf{Control group}} & & \multicolumn{2}{c}{\textbf{Treatment - Control}}  \\ \cline{3-5} \cline{7-8}
 & Total obs. & Obs. & Mean & SD & & Coefficient & \textit{p}-value \\\hline
\textit{Household Composition} &  &  &  &  &  & &  \\
Members resident & 960 & 259 & 4.908 & 1.739 &  & 0.001 & 0.995 \\
Number of adults (age 16 or older) & 960 & 259 & 3.162 &1.532
 &  & -0.007 & 0.970 \\
Number of female adults (age 16 or older) & 960 & 259 &1.838 & 1.025 &  &  -0.075 & 0.438 \\
Number of children (age 15 or younger) & 960 & 259 & 1.757 & 1.257 &  & 0.006 & 0.961  \\
Age of respondent & 960 & 259 & 40.919 & 9.358 &  & -1.304 & 0.183 \\
Respondent with no education & 960 & 259 & 0.849 & 0.358 &  & -0.028 & 0.439  \\
Respondent opted for vocational training & 960 & 259 & 0.216& 0.412 &  & -0.021  & 0.554  \\
Respondent married or living with partner & 960 & 259 &0.591 & 0.493 &  & 0.090 & 0.229 \\
Respondent is Buddhist & 960 & 259 & 0.757 & 0.430 &  & -0.028 & 0.570 \\
Respondent has Hahl etnicity & 960 & 259 & 0.656 & 0.476 &  & 0.101 & 0.519 \\
   &  &  &  &  &  & &  \\
\textit{Access to Credit} &  &  &  &  &  & & \\
Has borrowed from any source & 960 & 259 & 0.575 & 0.495 &  & 0.089 & 0.105  \\
Total amount of outstanding loans & 960 & 259 & \$473.475 & \$891.398 &  &  \$428.532 & 0.002  \\
Total expenditure on instalments last month & 960 & 259 & \$61.813 & \$125.061 &  & \$24.738 & 0.051  \\
   &  &  &  &  &  &  \\
\textit{Self-employment activities} &  &  &  &  &  & & \\
Any type of enterprise & 960 & 259 & 0.602 & 0.490 &  & -0.007 & 0.920 \\
Respondent has own enterprise & 960 & 259 & 0.378 & 0.486 &  & -0.009 & 0.896 \\
Expenses of any enterprise (past 12 months) & 960 & 259 & \$660.035 & \$1973.116 &  & \$48.614 & 0.793 \\
Revenues of respondent's enterprise (past 12 months) & 960 & 259 & \$777.497 & \$2065.149 &  & \$229.689  & 0.316 \\
Profit of respondent's enterprise (past 12 months) & 960 & 259 & \$243.740 & \$1753.716 &  & -\$3.919 & 0.960 \\
Total yearly hours worked (self and wage employment) & 960 & 259 & 84.124 & 79.564 & & -0.608 & 0.951 \\
    &  &  &  &  &  &  & \\
 \textit{Other employment activities} &  &  &  &  &  &  &\\
Number of income sources & 960 & 259 & 0.622 & 0.851 &  &-0.061 & 0.562  \\
Agriculture as income source & 960 & 259 & 0.120 & 0.325 &  & -0.023 & 0.561 \\
Private business as income source & 960 & 259 & 0.077 & 0.267 &  & 0.028 & 0.190  \\
Teaching as income source & 960 & 259 & 0.093 & 0.291 &  & 0.009 & 0.759 \\
Mining as income source & 960 & 259 & 0.012 & 0.107 &  & 0.013 & 0.276  \\
Any other income source & 960 & 259 & 0.243 & 0.430 &  & -0.083 &  0.094\\
    &  &  &  &  &  &  & \\
\textit{Consumption and savings} &  &  &  &  &  &  &\\
Total monthly expenditures on consumption & 960 & 259 & \$592.442 & \$922.301 &  & -\$35.242 & 0.606 \\
Total monthly expenditures on temptation goods & 960 & 259 & \$14.558 & \$28.237 &  & \$3.363 & 0.268  \\
Educational expenses last month & 960 & 259 & \$37.831 & \$96.361 &  & -\$4.942 & 0.482  \\
Total household savings & 960 & 259 & \$98.767  & \$314.582 &  & \$19.314 & 0.502  \\
    &  &  &  &  &  &  & \\
 \textit{Housing} &  &  &  &  &  &  &\\
Household owns a house & 960 & 259 & 0.390 & 0.489 &  & -0.078 & 0.319 \\
Household owns separate land & 960 & 259 & 0.367 & 0.483 &  & 0.100 & 0.157  \\
Type of dwelling$^{a)}$ & 960 & 259 &  1.363 & 0.542 &  & 0.066 & 0.385  \\
Household owns or rents other dwelling & 960 & 259 &  0.506& 0.501 &  & -0.046 & 0.541  \\
Household owns large household appliances & 960 & 259 & 0.529 & 0.500 &  & -0.005  & 0.929 \\
Household owns small household appliances & 960 & 259 & 0.776 & 0.418 &  & 0.010 & 0.810 \\
Distance to province center (km) & 960 & 259 & 113.321 & 52.360 &  & -0.985 & 0.959  \\
   &  &  &  &  &  &  & \\
\textit{Other} &  &  &  &  &  &  &\\
Number of cattle, sheep etc. owned by household & 960 & 259 & 39.614 & 52.705 &  &8.582  & 0.235  \\
Number of other animals owned by household & 960 & 259 & 0.243 & 0.703 &  & 0.216 & 0.077 \\
Number of livestock in village & 960 & 259  & 131180.800 & 49372.390 &  & 35388.340 & 0.075 \\
Number of people in village & 960 & 259 & 3609.351 & 3609.351 &  & 283.450 &  0.464 \\
Number of people in village center & 960 & 259 & 1008.529 & 418.624 &  & 104.661 & 0.562 \\
Number of families in village & 960 & 259 & 1002.915 & 312.139 &  &  101.908 &  0.418 \\
Number of doctors in village  & 960 & 259 & 5.942 & 3.578 &   & -1.310 & 0.311 \\
   &  &  &  &  &  & &  \\ \hline \hline
\end{tabularx}}}
\begin{tablenotes}[para]
\scriptsize{}	
 Note: All variables are recorded at baseline (pre-treatment) and all monetary variables are measured in \textsc{usd} (\$) \textsc{ppp} indexed to 2009. In addition 5 province dummies are included to capture effects at the aggregate level. A dummy indicating whether individuals were assigned group or individual treatment as well as dummies indicating follow-up dates are also included. Following the authors of the original study we derive the coefficient and $p$-value by regressing the respective baseline covariate on the treatment dummy and using standard errors clustered at the village level.\newline
 $^{a)}$Categorical variable: $1=Ger$, $2=House$, $3=Apartment$ and $4=Other$.\newline
\end{tablenotes}
\onehalfspacing
\end{center}
\restoregeometry
\newpage

\newgeometry{top=2cm, right=2cm ,left=1.7 cm, bottom=0.5cm}
\begin{center}
\singlespacing
\captionof{table}{Baseline covariates used for detecting heterogeneity for the RCT in Bosnia \& Herzegovina}\label{Bosnia}
\hspace{0 cm}
\hbox{
\scalebox{0.75}{
\begin{tabularx}{1.34\textwidth}{lccccccc} \hline \hline \\
 &  & \multicolumn{3}{c}{\textbf{Control group}} & & \multicolumn{2}{c}{\textbf{Treatment - Control}}  \\ \cline{3-5} \cline{7-8}
 & Total obs. & Obs. & Mean & SD & & Coefficient & \textit{p}-value \\\hline
\textit{Respondent Characteristics} &  &  &  &  &  & &  \\
Age of respondent & 994 & 443 & 37.097 & 11.954 &  & 1.290 & 0.092 \\
Female respondent & 995 & 444 & 0.405 & 0.492 &  & 0.008 & 0.789  \\
Respondent has never married & 995 & 444 & 0.236 & 0.425 &  & -0.008 & 0.772  \\
Respondent is married & 995 & 444 & 0.619 & 0.486 &  & -0.008 & 0.803  \\
Respondent is living with partner & 995 & 444 &0.009  & 0.095 &  &-0.005 & 0.298  \\
Respondent is separated  & 995 & 444 & 0.059 & 0.235 &  &0.014 & 0.372  \\
Respondent is widowed & 995 & 444 & 0.077 & 0.266 &  &0.005 &0.767  \\
Respondent is employed & 995 & 444 & 0.568 & 0.496 &  & -0.007& 0.831 \\
Respondent is unemployed & 995 & 444 & 0.261 & 0.440 & & 0.006 & 0.844  \\
Respondent is studying & 995 & 444 & 0.023 & 0.149 &  & -0.013 & 0.098  \\
Respondent is retired & 995 & 444 & 0.092 & 0.290 &  & 0.004 & 0.837  \\
Respondent has no education & 995 & 444 & 0.315 & 0.465 &  & 0.031 & 0.296 \\
   &  &  &  &  &  & &  \\
\textit{Household Characteristics} &  &  &  &  &  & &  \\
Number of female household members & 995 & 444 & 1.712 & 1.007 &  & 0.009 & 0.890 \\
Number of male household members & 995 & 444 & 1.736 & 0.962 &  & 0.169 & 0.009 \\
Number of adults (age 16 or older) & 995 & 444 & 2.543 & 1.045 &  & 0.147 & 0.032 \\
Number of children (age 15 or younger) & 995 & 444 & 0.842 & 1.024 &  & 0.045 & 0.494  \\
Number of household members attending school & 995 & 444 & 0.723 & 0.937 &  & 0.146 & 0.017 \\
Number of household members employed & 995 & 444 & 1.097 & 0.915 &  & 0.072 & 0.221 \\
Number of household members unemployed & 995 & 444 & 0.685 & 0.884 & & 0.021 & 0.710 \\
Number of household members retired & 995 & 444 & 0.313 & 0.515 &  &-0.001 & 0.979  \\
Number of household female members employed & 995 & 444 & 0.336 & 0.518 &  & 0.036 & 0.286 \\
   &  &  &  &  &  &  \\
\textit{Access to Credit} &  &  &  &  &  & & \\
Any type of loan & 995 & 444 & 0.583 & 0.494 &  & -0.019 & 0.549 \\
Number of loans & 995 & 444 & 0.802 & 0.864 &  &-0.030 & 0.582 \\
Total amount of outstanding loans & 995 & 444 & \$4611.661 & \$9625.364 &  & -\$1015.933 & 0.072 \\
   &  &  &  &  &  &  \\
\textit{Self-employment activities} &  &  &  &  &  & & \\
Any income from self-employment activity & 995 & 444 & 0.770 & 0.421 &  &0.001 & 0.969  \\
Respondent owns a business & 995 & 444 & 0.613 & 0.488  &  & 0.019 & 0.540  \\
Respondent owns a secondary business & 995 & 444 & 0.081 & 0.273  &  & 0.022 & 0.223  \\
Income from self-employment activity & 995 & 444 & \$7935.749 & \$11952.370  &  & \$269.835 & 0.745  \\
Income from agricultural activity & 995 & 444 & \$413.517 & \$1610.300  &  & -\$104.851 & 0.279  \\
    &  &  &  &  &  &  & \\
 \textit{Consumption and savings} &  &  &  &  &  &  &\\
Total consumption of food last week & 995 & 444 & \$118.397 & \$93.179 &  & -\$0.045 & 0.994  \\
Respondent eats higher nutritional diet than peers$^{a)}$  & 995 & 444 & 2.374 & 0.708  &  & 0.005 & 0.906 \\
Savings & 995 & 444 & \$1243.773 & \$3244.363 &  & \$14.101 & 0.944  \\
    &  &  &  &  &  &  & \\
 \textit{Housing} &  &  &  &  &  &  &\\
Respondent owns primary dwelling & 995 & 444 & 0.858 & 0.349 &  &0.035 & 0.101  \\
    &  &  &  &  &  &  & \\
 \textit{Other} &  &  &  &  &  &  &\\
Respondent thinks last year was financially successful$^{a)}$ & 995 & 444 & 2.689 & 0.614 &  & 0.035& 0.361  \\
Respondent thinks next year will be financially successful$^{a)}$ & 995 & 444 & 2.872 & 0.367  &  & 0.000 & 0.983  \\
   &  &  &  &  &  & &  \\ \hline \hline
\end{tabularx}}}
\begin{tablenotes}[para]
\scriptsize
 Note: All variables are recorded at baseline (pre-treatment) and all monetary variables are measured in \textsc{usd} (\$) \textsc{ppp} indexed to 2009.  Following the authors of the original study we derive the coefficient and $p$-value by regressing the respective baseline covariate on the treatment dummy and using standard errors robust to heteroskedasticity.\newline
 $^{a)}$Categorical variable:  $1=Disagree$, $2=Neutral$, and $3=Strongly$ $Agree$.\newline
\end{tablenotes}
\onehalfspacing
\end{center}
\restoregeometry

\newpage
\begin{center}
\captionof{table}{Performance comparison of the Elastic Net and Random Forest}\label{MLmethods} \
\hspace{0 cm}
\scalebox{0.83}{
\begin{tabularx} {1.13\textwidth}{lcc|cc|cc} \hline \hline \\
 & \multicolumn{2}{c|}{\textbf{Morocco}} & \multicolumn{2}{c|}{\textbf{Mongolia}} & \multicolumn{2}{c}{\textbf{Bosnia \& Herzegovina}} \\\cmidrule{2-7}
 & Best BLP & Best GATES  & Best BLP  & Best GATES  &  Best BLP  & Best GATES  \\
  & $\Lambda$ & $\bar{\Lambda}$ & $\Lambda$ & $\bar{\Lambda}$ & $\Lambda$ & $\bar{\Lambda}$
 \\\cmidrule{2-7}
 \textbf{Profit}  & & & & & & \\
Elastic Net & 105.035$^*$ & 8680.990$^*$ & 2.568$^*$ & 28.927$^*$ & 58.394 & 42957.210$^*$
\\
Random Forest & 65.109 & 4793.797 & 0.936 & 10.935 & 68.511$^*$ & 36946.750
\\ \cmidrule{2-7}
 \textbf{Consumption}  & & & & & &  \\
Elastic Net & 22.810 & 1228.565$^*$ & 225.227$^*$ & 15365.680 & 64.229$^*$ & 9996.229
\\
Random Forest & 29.820$^*$ & 1063.052 & 194.757 & 30605.060$^*$ & 51.640& 13882.880$^*$\\
& & & & & &\\
\hline \hline
\end{tabularx}}
\begin{tablenotes}[para]
\scriptsize
 Notes$:$
 The table presents estimates of $\Lambda$ and $\bar{\Lambda}$ which measures how well the Elastic Net and Random Forest algorithms fit the Best Linear Predictor (BLP) and GATES functions for profits and consumption, respectively, in a given RCT. The estimates are derived from equation (\ref{lambda_B})
 and (\ref{lambda_G}) and given by the medians over 50 splits. Profits and consumption are monthly profits from business activities and total monthly consumption, both measured in USD PPP indexed to 2009. The asterisk indicates the better performing machine learning method for each category (e.g. for a particular measure (BLP or GATES) for a particular outcome (profits or consumption) in a particular RCT (Morocco, Mongolia or Bosnia \& Herzegovina)).
\end{tablenotes}
\onehalfspacing
\end{center}

\begin{center}
\captionof{table}{Predictive Power of Covariates for Treatment Effect Heterogeneity in Profits}\label{HHvsAGG}
\hspace{0 cm}
\hbox{
\scalebox{0.9}{
\begin{tabularx} {0.6\textwidth}{lcc} \hline \hline \\
 & \textbf{Morocco} & \textbf{Mongolia} \\ \cmidrule{2-3}
  Aggregate level covariates & 0.87 & 0.67 \\
 Household level covariates & 0.49 & 0.81 \\
 All covariates & 0.94 & 0.91
 \\  \hline \hline
\end{tabularx}}}
\begin{tablenotes}[para]
\scriptsize
 Notes$:$
The table presents the medians over 50 splits of the adjusted $R^2$ values from an regression of an indicator dummy that equals one if an individual belongs to the top 25 pct. of people who experienced the highest treatment ($G_4$) effects on profits and zero if an individual belongs to the bottom 25 pct. ($G_1$) defined by the Elastic Net proxy predictor $S(Z)$. The first row show the adjusted $R^2$ from the regression that controls for aggregate (village/province) level covariates and strata dummies only. The second row shows the adjusted $R^2$ from the regression that controls for household level covariates only. Finally, the last row shows the adjust $R^2$ from the regression controlling for all covariates (i.e. both aggregate and household level). The exact household covariates and aggregate covariates / strata dummies used in the RCTs are available in Table \ref{Morocco} and \ref{Mongolia}.
\end{tablenotes}
\end{center}

\newpage

\begin{center}
\captionof{table}{ATE and HET estimates for profits and consumption across RCTs (Random Forest)}\label{RFappend1}
\hspace{0 cm}
\hbox{
\scalebox{0.75}{
\begin{tabularx} {1.33\textwidth}{lcc|cc|cc} \hline \hline \\
 & \multicolumn{2}{c}{\textbf{Morocco}} & \multicolumn{2}{c}{\textbf{Mongolia}} & \multicolumn{2}{c}{\textbf{Bosnia \& Herzegovina}} \\
 & ATE ($\hat{\beta}_1$) & HET ($\hat{\beta}_2$) & ATE ($\hat{\beta}_1$) & HET ($\hat{\beta}_2$) & ATE ($\hat{\beta}_1$) & HET ($\hat{\beta}_2$) \\ \cmidrule{2-7}
Profits & 27.159 & 0.135 & -1.567 & 0.204 & 76.812 & 0.107
\\
& (-23.107,80.741) & (-0.022,0.305) & (-3.780,0.676) & (-0.213,0.669) & (-70.168,208.003) & (-0.420,0.639)
 \\
 & [0.581] & [0.186] & [0.305] & [0.632] & [0.615] & [0.926]\\
Consumption & -10.036 & 0.062
 & 129.373 & 0.517 & -29.249 & 0.069 \\
& (-32.451,14.143) & (-0.110,0.405) & (-29.102,278.685) & (-0.056,0.921) & (-180.348,114.931) & (-0.323,0.500) \\
& [0.694] & [0.451] & [0.224] & [0.153] & [1.000] & [1.000] \\
& \multicolumn{6}{c}{} \\  \hline \hline
\end{tabularx}}}
\begin{tablenotes}[para]
\scriptsize
 Notes$:$
The table presents estimates of the average treatment effects (ATE) and a measure of heterogeneity in treatment effects (HET) for profits and consumption. The estimates are given by the median of the estimates of $\beta_1$ and $\beta_2$ obtained by applying WLS to equation (\ref{BLP}) over 50 splits. Similarly, the sample splitting adjusted confidence intervals and $p$-values are obtained by the median over 50 splits with the $p$-values being twice this value to correct for sample splitting.  Clustered standard errors at the village level for Morocco and Mongolia, and robust standard errors for Bosnia \& Herzegovina are used. Strata dummies at the village level for Morocco and at the province level for Mongolia are included. The proxies S(Z) and B(Z) that enters equation (\ref{BLP}) are estimated using an Random Forest on the covariates presented in Tables \ref{Morocco}, \ref{Mongolia}, and \ref{Bosnia}, for Morocco, Mongolia and Bosnia \& Herzegovina respectively. Profits and consumption are monthly profits from business activities and total monthly consumption, both measured in USD PPP indexed to 2009. \newline
 $\bullet$ 90\% confidence intervals in parentheses.\newline
 $\bullet$ P-values for the hypothesis that the parameter is equal to zero in brackets.\newline
\end{tablenotes}
\end{center}

\newpage

\restoregeometry
\newgeometry{top=3cm, bottom=3cm, left=2cm, right=2cm}
\begin{center}
\captionof{table}{GATES estimates for profits and consumption across RCTs (Random Forest)}\label{RFappend2}
\hspace{0.5 cm}
\hbox{
\scalebox{0.9}{
\begin{tabularx} {0.9\textwidth}{lccc} \hline \hline
& \multicolumn{3}{c}{\textbf{Panel A: Morocco}} \\
 & 25 \% most ($\hat{\gamma}_{4}$) & 25 \% least ($\hat{\gamma}_1$) & Difference ($\hat{\gamma}_{4}$ - $\hat{\gamma}_{1}$) \\\cmidrule{2-4}
Profits & 144.754
 & -6.117 & 149.716  \\
& (11.225,295.095) & (-113.139,109.946) & (-46.900,335.198) \\
& [0.070]& [1.000]& [0.305] \\
Consumption & 0.167 & -47.423 & 45.517  \\
& (-54.352,57.458) & (-114.765,19.014) & (-51.346,146.552) \\
& [0.866] & [0.308] & [0.729] \\\cmidrule{2-4}
& \multicolumn{3}{c}{\textbf{Panel C: Mongolia}} \\
 & 25 \% most ($\hat{\gamma}_{4}$) & 25 \% least ($\hat{\gamma}_1$) & Difference ($\hat{\gamma}_{4}$ - $\hat{\gamma}_{1}$) \\\cmidrule{2-4}
Profits & -0.200 & -3.284 & 3.079 \\
& (-4.486,3.808) & (-7.839,1.281) &  (-2.708,9.138) \\
& [1.000] & [0.323] & [0.569]
 \\
Consumption & 377.988 & 75.083 & 298.610\\
& (-92.411,850.317) & (-101.806,253.215) & (-204.444,811.183) \\
& [0.239] & [0.817] & [0.439] \\\cmidrule{2-4}
& \multicolumn{3}{c}{\textbf{Panel D: Bosnia}} \\
 & 25 \% most ($\hat{\gamma}_{4}$) & 25 \% least ($\hat{\gamma}_1$) & Difference ($\hat{\gamma}_{4}$ - $\hat{\gamma}_{1}$) \\\cmidrule{2-4}
Profits & 133.482 & 20.853 & 126.804 \\
& (-158.902,432.404) & (-296.607,298.607) &  (-340.747,567.488) \\
& [0.701] & [0.885] & [0.887] \\
Consumption & 5.265 & -31.944 & -1.524\\
& (-265.275,291.638) & (-343.649,298.058) & (-425.770,447.089) \\
& [0.986] & [1.000] & [1.000]\\ \hline \hline
\end{tabularx}}}
\begin{tablenotes}[para]
\scriptsize	
Notes$:$ The table presents estimates of the Sorted Group Average Treatment Effects (GATES) for monthly profits and monthly consumption measured in \textsc{usd (ppp)} indexed to 2009. Panel A, B and C shows the estimates for the RCTs in Morocco, Mongolia and Bosnia \& Herzegovina respectively. The final estimates, confidence intervals and $p$-values are found by estimating (\ref{GATES}) by WLS over 50 splits and then taking the median of the split-dependent estimates. Clustered standard errors at the village level for Morocco and Mongolia, and robust standard errors for Bosnia \& Herzegovina are used. Strata dummies at the village level for Morocco and at the province level for Mongolia are included. The two non-overlapping groups G1 and G4 are defined by the proxy $S(Z)$ which is estimated using a Random Forest. G4 consists of the 25 pct. most affected and G1 of the 25 pct. least  affected. The difference is found by testing whether the group estimates are equal. The final estimate, confidence interval and $p$-value for this quantity are found in the same way as the estimates of $\gamma_1$ and $\gamma_4$. \newline
 $\bullet$ 90\% confidence intervals in parentheses.\newline
 $\bullet$ $p$-values for the hypothesis that the parameter is equal to zero in brackets.\newline
\end{tablenotes}
\onehalfspacing
\end{center}

\end{document}